\documentclass[sigconf]{acmart}

 \copyrightyear{2024} 
\acmYear{2024} 
\setcopyright{acmlicensed}\acmConference[CCS '24]{Proceedings of the 2024
 ACM SIGSAC Conference on Computer and Communications Security}{October
 14--18, 2024}{Salt Lake City, UT, USA}
 \acmBooktitle{Proceedings of the 2024 ACM SIGSAC Conference on Computer
 and Communications Security (CCS '24), October 14--18, 2024, Salt Lake City,
 UT, USA}
 \acmDOI{10.1145/3658644.3690285}
 \acmISBN{979-8-4007-0636-3/24/10}

\usepackage{xcolor}
\usepackage[normalem]{ulem}
\usepackage{circledsteps}
\usepackage{algorithm}
\usepackage{algpseudocode}
\usepackage{subfig}
\usepackage{xspace}
\usepackage{fontawesome}
\usepackage{pifont}
\usepackage{array}
\usepackage{circledsteps}
\usepackage[skins]{tcolorbox}

\newcommand{\system}{\textsc{GAZEploit}\xspace}
\newcommand{\hliwona}[1]{{\fontfamily{kurier}\selectfont \textbf{#1}}}

\begin{document}

\title[\system]{\system: Remote Keystroke Inference Attack by Gaze Estimation from Avatar Views in VR/MR Devices}

\author{Hanqiu Wang}
\affiliation{%
  \institution{Department of Electrical and Computer Engineering, University of Florida,}
  \city{Gainesville}
  \state{FL}
  \country{USA}
  }
\email{wanghanqiu@ufl.edu}

\author{Zihao Zhan}
\affiliation{%
  \institution{Department of Computer Science, Texas Tech University,}
  \city{Lubbock}
  \state{TX}
  \country{USA}
  }
\email{zihao.zhan@ttu.edu}

\author{Haoqi Shan}
\affiliation{%
  \institution{CertiK,}
  \city{New York}
  \state{NY}
  \country{USA}
  }
\email{haoqi.shan@certik.com}

\author{Siqi Dai}
\affiliation{%
  \institution{Department of Electrical and Computer Engineering, University of Florida,}
  \city{Gainesville}
  \state{FL}
  \country{USA}
  }
\email{dais@ufl.edu}

\author{Maximilian Panoff}
\affiliation{%
  \institution{Department of Electrical and Computer Engineering, University of Florida,}
  \city{Gainesville}
  \state{FL}
  \country{USA}
  }
\email{m.panoff@ufl.edu}

\author{Shuo Wang}
\affiliation{%
  \institution{Department of Electrical and Computer Engineering, University of Florida,}
  \city{Gainesville}
  \state{FL}
  \country{USA}
  }
\email{shuo.wang@ece.ufl.edu}

\begin{abstract}

    The advent and growing popularity of Virtual Reality (VR) and Mixed Reality
    (MR) solutions have revolutionized the way we interact with digital
    platforms. The cutting-edge gaze-controlled typing methods, now
    prevalent in high-end models of these devices, e.g., Apple Vision Pro, have
    not only improved user experience but also mitigated traditional keystroke
    inference attacks that relied on hand gestures, head movements and acoustic
    side-channels. However, this advancement has paradoxically given birth to a
    new, potentially more insidious cyber threat, \textbf{\system}.

    In this paper, we unveil \textbf{\system}, a novel eye-tracking based attack
    specifically designed to exploit these eye-tracking information by leveraging
    the common use of virtual appearances in VR applications. This widespread
    usage significantly enhances the practicality and feasibility of our attack
    compared to existing methods. \textbf{\system} takes advantage of this
    vulnerability to remotely extract gaze estimations and steal sensitive
    keystroke information across various typing scenarios—including messages,
    passwords, URLs, emails, and passcodes. Our research, involving 30
    participants, achieved over 80\% accuracy in keystroke inference.
    Alarmingly, our study also identified over 15 top-rated apps in the Apple
    Store as vulnerable to the \textbf{\system} attack, emphasizing the urgent
    need for bolstered security measures for this state-of-the-art VR/MR text
    entry method.

\end{abstract}



\maketitle

\section{Introduction}
\label{sec:intro}

The proliferation of Virtual and Mixed Reality (VR/MR) headsets in the consumer
market is a recent phenomenon, despite the existence of the technology for many
years. With an impressive user base of 171 million devices and
counting~\cite{Katatikarn_2022}, VR/MR technology is quickly breaking free from
its initial entertainment confines. Today, it is being utilized in a wide array
of fields, including education, professional training, social interaction, and
remote work environments. Given that text entry is a fundamental interaction
with electronic devices, VR devices commonly incorporate virtual keyboards to
facilitate efficient typing tasks.

The evolution of typing methods on VR's virtual keyboards is noteworthy. Earlier
VR devices, such as the HTC Vive and Oculus Rift, employed
controllers~\cite{controlleroverview} as optical pointers for key selection on a
virtual keyboard~\cite{htccontroller}. More recent devices like the Meta Quest
have transitioned to using external cameras to capture users' hand gestures for
typing. In the latest VR devices, eye-tracking technology has gained prominence,
enabling the potential implementation of gaze-controlled typing, an intuitive,
efficient, and user-friendly text entry method. Major manufacturers, including
Microsoft, Meta, and HTC, have acknowledged the potential of this method. Recent
research~\cite{rajanna2018gaze, eyetrackingkeyboard2022, zhao2023gaze} has
further confirmed its feasibility. Notably, Apple's latest VR device, the Apple
Vision Pro MR headset, has integrated gaze-controlled typing as the primary text
entry method, marking a significant shift in the industry~\cite{visionPro}.

Given that text entry often involves confidential data such as passwords, PINs,
and private messages, recent researches have identified vulnerabilities
associated with different virtual keyboard typing methods. Prior
studies~\cite{hiddenReality,gearVR,luo2024eavesdropping,wifiVR,air-tapping,goingMotions}
have demonstrated the ability to recover keystrokes from side-channel
information, captured by motion sensors on the headset or through external
optical and acoustic recordings of users. These investigations focused on
controller-based, head-pose-based, and hand-gesture-based typing methods. The
approach taken by Apple's Vision Pro, which utilizes an eye tracker as the
pointer, minimizes the need for hand and controller movements as well as head
movements, making many traditional attack vectors ineffective.

While the integration of gaze-controlled typing methods in Apple Vision Pro has
significantly enhanced user interaction, it also introduces new security and
privacy risks. Apple recognizes the sensitivity of eye-tracking data and has
implemented stringent measures to confine its access~\cite{apple_privacy_2024}.
This data is processed exclusively at the system level, with applications only
receiving the final computation results, not the underlying camera data or
information used to generate them. Moreover, Apple's security measures stipulate
that the virtual keyboard is managed only by visionOS~\cite{visionOS} and cannot
be accessed or controlled by any app. Despite these precautions, our research
uncovers a significant vulnerability associated with the virtual avatar
technologies like Apple's Persona. These technologies, while enhancing digital
communication, can inadvertently expose critical facial biometrics, including
eye-tracking data, through video calls where the user's virtual avatar mirrors
their eye movements. Leveraging this vulnerability, we developed \system, a
novel attack that can infer eye-related biometrics from the avatar image to
reconstruct text entered via gaze-controlled typing. This finding shows that
even high-level abstractions provided to applications and users can be
reverse-engineered to reveal sensitive data at the system level, effectively
bypassing Apple's security measures such as the app-inaccessible virtual
keyboard and eye-tracking data.

\begin{figure}[htbp]
    \centering
    \subfloat[]{\includegraphics[height=.5\columnwidth]{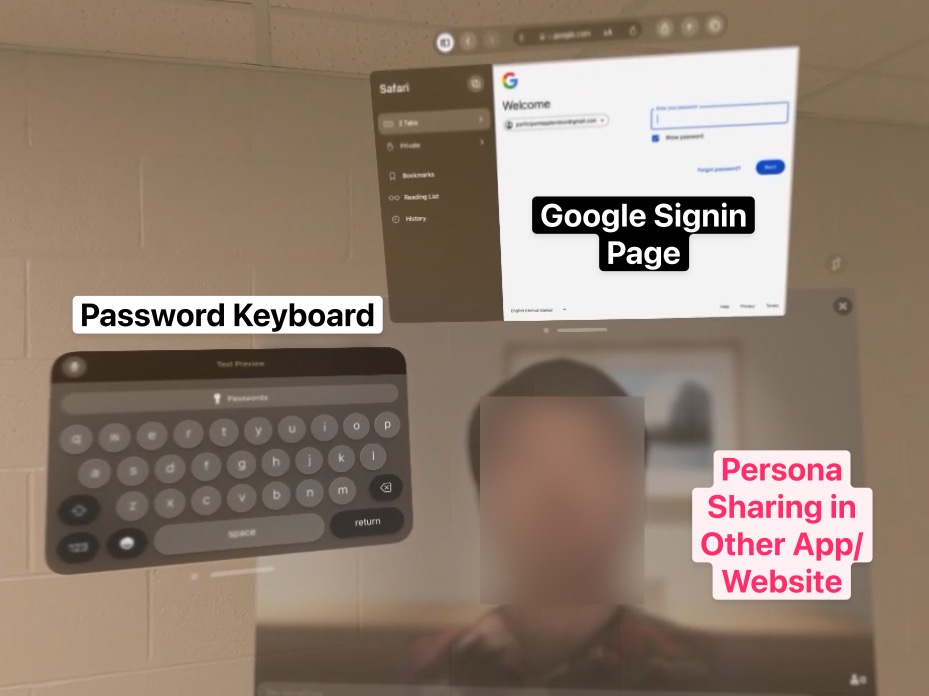}}
    \hfill
    \subfloat[]{\includegraphics[height=.5\columnwidth]{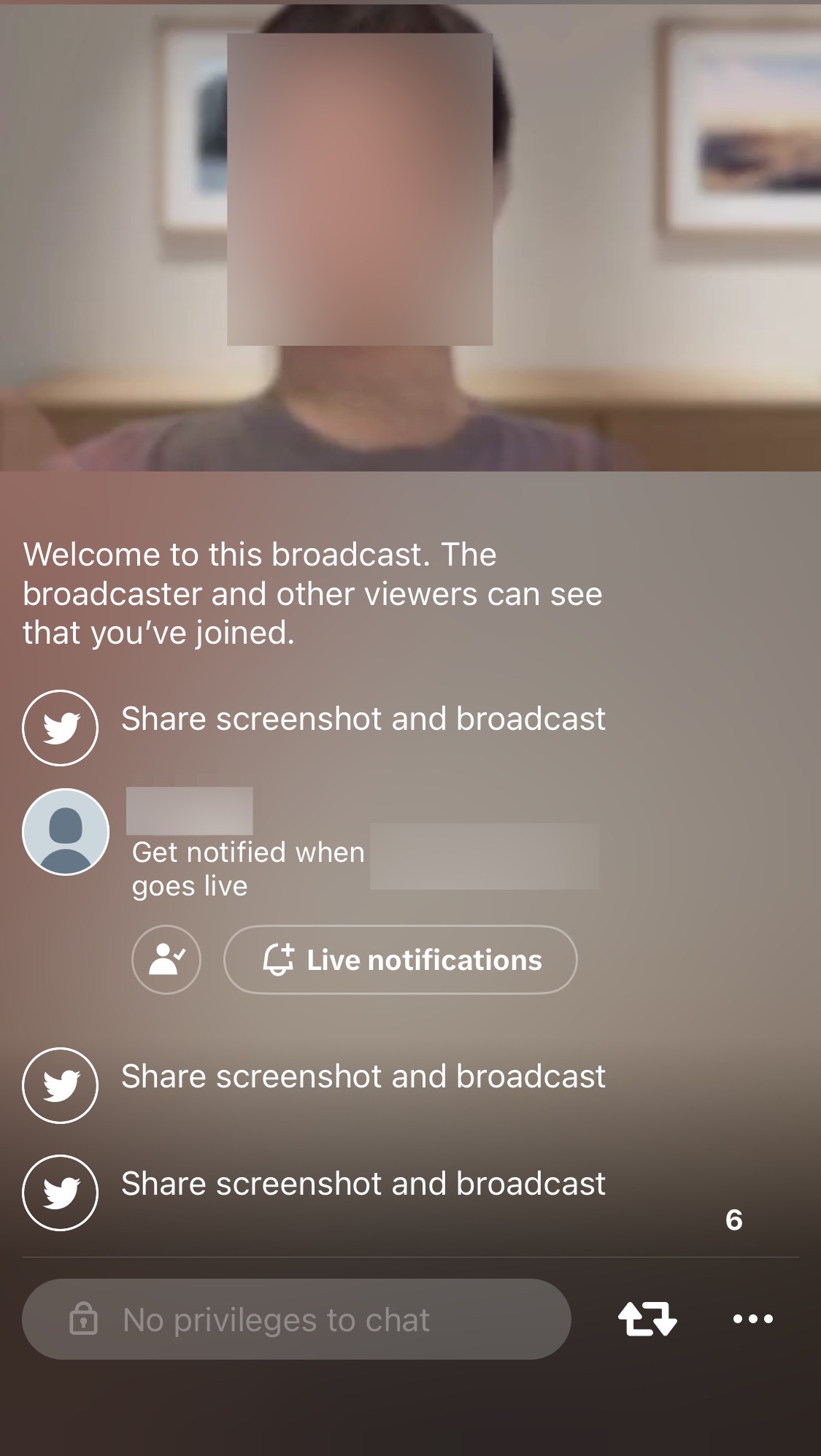}}
    \caption{(a) A victim is logging into a Google account while sharing the
        Persona. (b) An attacker viewing the victim's Persona from a live video on
        X. }
    \label{fig:intro}
\end{figure}

The \system attack leverages the vulnerability inherent in gaze-controlled text
entry when users share a virtual avatar, as illustrated in
Figure~\ref{fig:intro}. Virtual avatars, whether shared through video calls,
online meeting apps, live streaming platforms, or potentially malicious
websites, pose a significant privacy risk by potentially exposing user
information such as login credentials. By remotely capturing and analyzing the
virtual avatar video, an attacker can reconstruct the typed keys. Notably, the
\system attack is the first known attack in this domain that exploits leaked
gaze information to remotely perform keystroke inference.

To evaluate the effectiveness of our \system attack, we conducted experiments
with 30 participants. This involved collecting data from videos recorded during
Zoom meetings in two sessions.
The data collection was designed to mirror
real-world user behavior closely, with minimal constraints imposed on
participants. Our evaluation demonstrated that our attack model achieved more
than 85.9\% precision and 96.8\% recall in identifying click candidates within the
video clips \textbf{without any prior knowledge of victims' personal typing
    habits nor their typing speed}. Key findings from the top-5 keystroke inference
accuracy include: 92.1\% for messages, 77.0\% for passwords, 73.0\% for
passcodes (PIN), and 86.1\% for email, URL, and webpage text entry.

To summarize, the main contributions of this paper are:
\begin{itemize}

    \item We analyze popular VR/MR devices, identifying two new attack
          scenarios and associated attack vectors exploiting eye-tracking
          systems in benign applications such as conference apps and public broadcasts.
    \item We develop the \system attack, the first known method exploiting the
          vulnerabilities discovered in our research to perform remote keystroke
          inference on VR/MR devices.
    \item We evaluate the \system attack using a dataset collected from 30
          participants in real-world VR settings demonstrating over 80\% precision in
          top 5 keystroke guesses.
    \item We present end-to-end attack scenarios \footnote{Readers can view our
              practical attack scenarios and associated video clips by visiting
              \url{https://sites.google.com/view/Gazeploit/}.} and discuss the
          security implications of our findings. We also propose potential
          countermeasures for both VR device manufacturers and developers.

\end{itemize}

The organization of the paper is as follows: Section~\ref{sec:back} provides an
overview of the VR device, its key features, and the gaze estimation techniques
utilized in this attack. Section~\ref{sec:threat} elaborates on the identified
vulnerabilities and the associated threats. Section~\ref{sec:attack} offers a
comprehensive demonstration of the preliminary attack, including detailed
discussions of the steps, challenges, and methods involved.
Section~\ref{sec:expe} describes our approach to data collection, while
Section~\ref{sec:eval} presents the performance and accuracy results of our
attack model. In Section~\ref{sec:disc}, we discuss potential countermeasures
and significant insights. Section~\ref{sec:relwork} reviews relevant literature,
and Section~\ref{sec:conclu} concludes the paper.

\section{Background}
\label{sec:back}

\subsection{Eye trackers for VR/MR and Text Entry}
\label{sec:2.1}

In this section, we explore the underlying technologies and features of modern
VR/MR devices that are pivotal to both enhancing user interaction and presenting
new cybersecurity challenges.

\noindent\textbf{VR Eye Trackers and Their Usage} Advanced VR/MR devices such as
the Apple Vision Pro~\cite{visionPro}, PSVR 2~\cite{psvr2}, and Meta Quest
Pro~\cite{metaquestprovr} integrate eye trackers into their headsets. This
technology operates by using cameras and sensors near the lenses to capture
detailed images of the eyes. These images, calibrated to each individual's
unique anatomical features, are then processed to generate eye-tracking data,
including metrics like pupil size, eye openness, and gaze direction. From these
basic measurements, complex features such as fixations, saccades, smooth
pursuits, and scanpaths can be derived~\cite{blascheck2017visualization}. For
further explanation, fixations represent the concentration of a user's gaze on a
specific area over a certain period, while saccades are the rapid eye movements
between these fixations. The sequence of these movements and fixations is known
as a scanpath, while smooth pursuits refer to the eye's movement while tracking
moving objects.

The integration of these features into VR/MR systems significantly enhances user
experience and system efficiency~\cite{qian2023reinforcement}. By enabling users to interact with and
navigate the virtual environment through eye movements, the technology provides
a more immersive experience. 
For instance, 
devices like the Meta Quest Pro allow avatars to
mimic realistic eye contact and facial expressions, enhancing the authenticity
of virtual conversations~\cite{metaQuestPro}. Similarly, in the Apple Vision
Pro, eye tracking is used to animate the avatar's eyes accurately, thereby
improving the realism of social interactions in virtual settings.

\noindent\textbf{Gaze-controlled Typing in VR} Building on the eye-tracking
capabilities discussed in the previous section, developers have proposed
gaze-assisted typing in VR devices~\cite{zhao2023gaze}. The Apple Vision Pro
represents a significant step in this direction, being the first VR device to
incorporate an eye tracker as the default keystroke entry method at the system
level. The Vision Pro offers two typing methods. The first employs eye gaze to
select keys, with a finger snap to confirm. The second method involves poking in mid-air at
the virtual keyboard using a single finger~\cite{appleTyping}. This differs from
air-tapping, which uses multiple fingers and is generally less preferred due to
its complexity and lower precision~\cite{Hiro2018airtap}. 
The effectiveness of gaze-controlled typing is also supported by participants' feedback, All favored it for its higher accuracy, the flexibility it offers in positioning the virtual keyboard, and the reduced need for extensive hand or body movements.

\subsection{Virtual Avatar and Virtual Camera}

Advanced VR systems often allow users to create digital avatars, which serve as
their virtual representations. These avatars vary in complexity, from simple
head-and-hand depictions to fully detailed human-like characters, depending on
the VR system and its purpose. For example, Meta VR devices have a
feature~\cite{metavereavatar} called ``Metaverse Avatar'' enabling users to
interact as 3D avatar in apps like Zoom and Horizon Workrooms. Similarly,
popular VR games like VRChat~\cite{vrchat} and Roblox~\cite{roblox} use avatars
to facilitate user interactions in virtual environments.

In the Apple Vision Pro, the Persona~\cite{roth_2023} feature takes this concept
further by displaying a dynamic representation of the user's face and hands,
capturing facial expressions and eye movements with high fidelity using the
integrated eye-tracker. Unlike other VR systems that use 3D avatars, Persona
operates more like a 2D camera view. This allows apps requiring camera access to
use Persona once granted permission by the user, even if they aren't
specifically designed for VR. Websites can also access Persona, extending its
usability beyond typical VR contexts. However, this flexibility also presents
additional security risks. As discussed earlier in Section~\ref{sec:intro},
VisionOS privacy standards protect sensitive eye-gaze data. Yet, developers may
overlook that this data could be exposed through the Persona view, creating a
vulnerability that our \system attack exploits.

\subsection{Appearance-based Eye Gaze Estimation}

Appearance-based eye gaze estimation is a method used in computer vision and
human-computer interaction that determines where a person is looking based
solely on images of their eye or face. 
Thus, it offers greater flexibility, but can also result in
reduced accuracy due to variables like lighting changes or head
movements~\cite{gazeReview}. To achieve this, modern gaze estimation techniques
frequently utilize deep learning architectures, with convolutional neural
networks (CNNs)~\cite{zhang2015appearance} being prevalent due to their strong
performance with image data. Additionally, other architectures such as recurrent
 neural networks (RNNs) and Transformers are also being investigated
~\cite{gazeReview}. The
inherent flexibility and minimal hardware requirements of appearance-based gaze
estimation make it particularly suited for environments where traditional
tracking devices are impractical. However, this also opens up significant
privacy concerns, as these methods can be exploited by attackers to extract
confidential gaze data~\cite{elfares2024privateyes}.

\section{Threat Model}
\label{sec:threat}
\subsection{Attack Scenarios}

Our threat model is based on a VR user who participates in a variety of
activities while sharing their virtual avatar, including multiple text entry
sessions via eye tracking. The user can freely arrange application windows
within the virtual space, place them in any direction at desired distances, and
select their own text content for input. We assume that \textbf{the attacker
    only needs a video recording of the user's virtual avatars during these
    activities, which can be obtained through various channels such as video conferencing, live
    streaming} that request access to the user's virtual
representation. The virtual avatars only contain users facial expressions and
eye gaze information. Hand gestures are not needed in the captured video. No
additional knowledge about the user's behaviors, such as the timing of text
entry sessions or the placement of windows/keyboards, is required. By analyzing
these video recordings, the attacker can infer when the user is typing and
recover the typed text content. We define two main attack scenarios under this
model:

\begin{enumerate}
    \item Streamers who publicly share their Persona views during a live stream,
          exposing their eye gaze data to potential attackers.
    \item Eye gaze data leaked via video conference apps, such as Zoom and
          Microsoft Teams, if a user shares the Persona virtual camera view with an
          attacker during a meeting or a call.
\end{enumerate}

The complexity, stealthiness, and frequency of the attack differ based on the
attack vectors. For streamers, no prerequisites are required for attackers,
making it the most straightforward scenario. However, the potential victims are
limited to those who stream their VR activities. In the case of attacks via
conference apps, the complexity and prevalence depend on the nature of the
online meetings. Highly confidential meetings may present more obstacles, while
public webinars, open to all attendees, may be easier to exploit.

\begin{figure}[htbp]
    \centering
    \begin{minipage}[c]{0.5\columnwidth}
        \centering
        \subfloat[]{\label{fig:vision_greendot}\includegraphics[width=\textwidth]{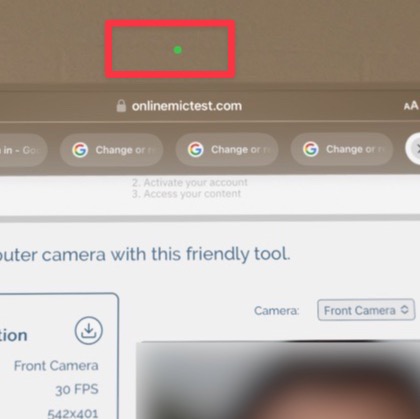}}
    \end{minipage}
    \hfill
    \begin{minipage}[c]{0.4\columnwidth}
        \centering
        \subfloat[]{\label{fig:iphone_greendot}\includegraphics[width=\textwidth]{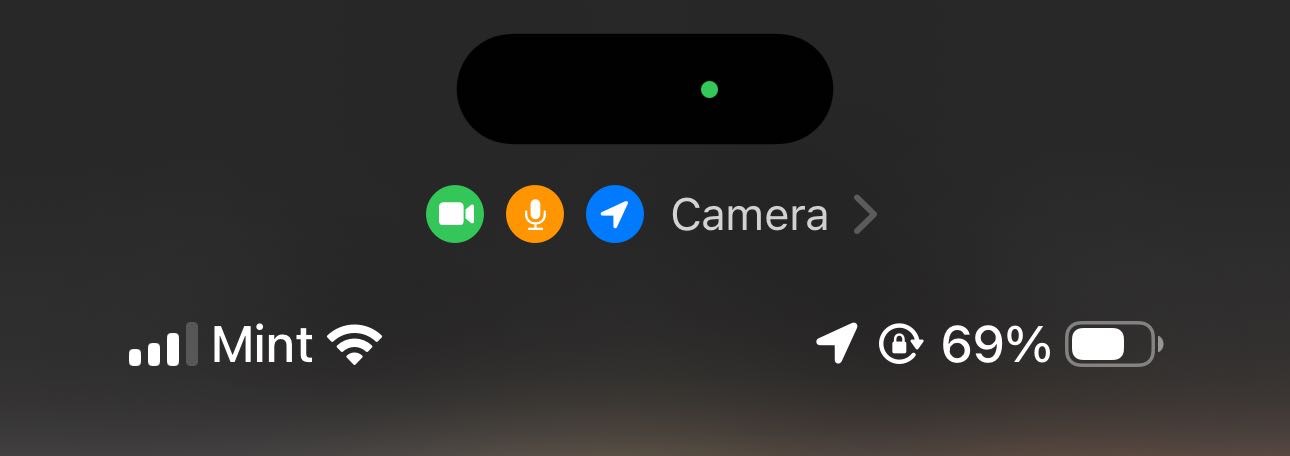}}\
        \subfloat[]{\label{fig:macos_greendot}\includegraphics[width=\textwidth]{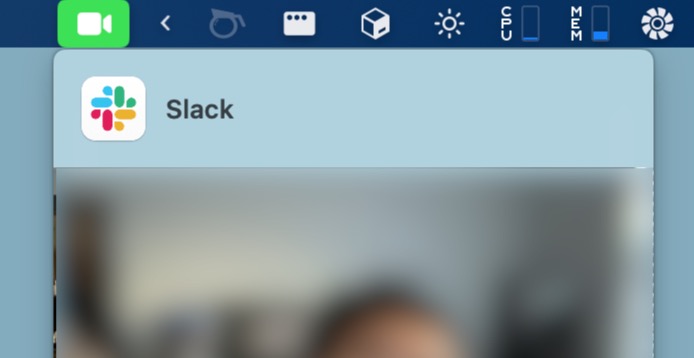}}
    \end{minipage}
    \caption{Camera access indicators on different platforms (a) Apple Vision Pro's subtle green dot that can be covered by any other window. (b) iPhone's camera access indicator. (c) MacBook's system menu bar camera indicator.}
    \label{fig:greendots}
\end{figure}

Moreover, 
\system can be exceptionally stealthy, with users potentially unaware their Persona is being recorded. Unlike laptops or smartphones that alert users with a hardware LED or system notification, the Apple Vision Pro only uses a small green dot to indicate virtual camera usage, as shown in Figure~\ref{fig:greendots}. This indicator can easily be \textbf{\textit{obscured}} if another app window overlays the Persona recording window, hiding the visual cue. The attacker's view remains unaffected. Additionally, the green dot is very small compared to the large VR/MR workspace, making it easy to miss. This subtle indicator, combined with the unique spatial configurations in a 3D VR workspace, enhances the attack's stealthiness, making it less noticeable than on traditional 2D devices.

\subsection{Attack Vectors on Various Typing Scenarios}
\label{sec:3.2}

Building upon the attack scenarios discussed earlier, we further divide the
attack into four distinct vectors. These vectors are based on different typing
scenarios in VR/MR environments. Each scenario has unique characteristics, such
as keyboard size and typing patterns, which influence the attacker's strategy
and the attack's effectiveness.

\begin{figure}[htbp]
    \centering
    \begin{minipage}[c]{0.3\columnwidth}
        \centering
        \subfloat[]{\label{fig:keyboard1}\includegraphics[width=\textwidth]{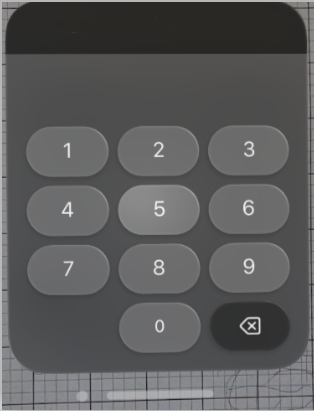}}
    \end{minipage}
    \hfill
    \begin{minipage}[c]{0.6\columnwidth}
        \centering
        \subfloat[]{\label{fig:keyboard2}\includegraphics[width=\textwidth]{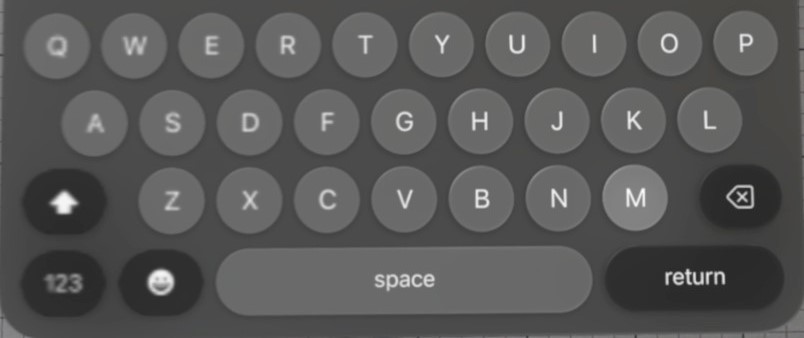}}\
        \subfloat[]{\label{fig:keyboard4}\includegraphics[width=\textwidth]{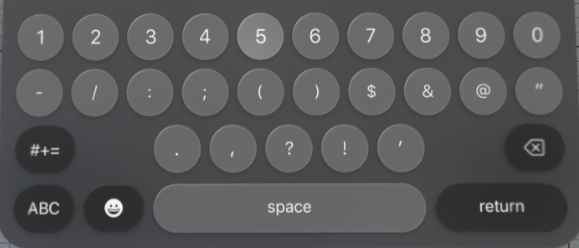}}
    \end{minipage}
    \caption{Keyboards used on Apple Vision Pro (a) Passcode(PIN) keyboard (b) Default QWERTY keyboard (c) Number and special character keyboard}
    \label{fig:keyboards}
\end{figure}

\noindent\textbf{Passcode (PIN) Inference:} The passcode (PIN) keyboard, as
shown in Figure~\ref{fig:keyboard1}, is exclusively numeric. While Apple Vision
Pro employs Optic ID~\cite{opticId} as an alternative to passcodes, this
keyboard remains integral for tasks such as device unlocking, phone number
entry, PIN input in financial apps, and two-factor authentication. The compact
design of this keyboard results in gaze angles focused within a narrow area,
providing a distinct pattern for the attacker to exploit. Additionally, the
typical input lengths—6 digits for PINs and passcodes, 10 for phone numbers—give
further clues about the nature of the data being entered.

\noindent\textbf{Password Inference:} The entry of passwords is unique in that
it often requires toggling between numeric and alphabetic segments of the
keyboard. This may involve using a ``\texttt{shift}'' or ``\texttt{num}'' key to
access special characters and numbers. Additionally, passwords usually do not
include spaces. These features can serve as key indicators to identify password
entry sessions, distinguishing them from other types of text entry. The
keyboards used for password entry are shown in Figure~\ref{fig:keyboard2} and
Figure~\ref{fig:keyboard4}.

\noindent\textbf{URL and Email Address Input:} The typing of URLs or email
addresses presents unique patterns that can be exploited for inference attacks.
These inputs often contain specific sequences or symbols, such as
``\texttt{.com}'', ``\texttt{.edu}'', ``\texttt{.net}'', or ``\texttt{@}'',
which can signal the completion of a web address or email address. 
The keyboard
used for these inputs remains the QWERTY layout (shown in
Figure~\ref{fig:keyboard2}), but with the space key replaced by space,
``\texttt{@}'', and ``\texttt{.}'' keys.

\noindent\textbf{Word Inference for Message Recovery:} In scenarios where users
draft messages or articles outside of web browsers, they often switch between
the keyboards displayed in Figure~\ref{fig:keyboard2} and
Figure~\ref{fig:keyboard4}. These keyboards, in comparison to the passcode
keyboard, cover a broader range of gaze regions. Furthermore, the presence of
multiple spaces between words, which are absent in password, URL, or email
address inputs, can serve as an identifiable pattern.

\subsection{Limitations on Attack Scenarios}

Our attack can be executed as long as several conditions are met by the VR platform: \Circled{\textbf{1}} the VR model is equipped with an eye tracker, 
\Circled{\textbf{2}} the VR device can create sharable virtual avatars displaying users' eye movements, and \Circled{\textbf{3}} the VR device supports a gaze-controlled keystroke input method.

While many VR devices have already met the first two requirements, few of them met requirement \Circled{\textbf{3}}. Currently, only the Apple Vision Pro satisfies all three requirements, which limits the applicability of \system. This limitation is not due to any shortcomings of \system but rather the limited adoption of gaze-typing technology in VR devices currently available in marketplace. 

However, gaze-typing is an emerging technology that is expected to become more popular in consumer devices. Companies like Meta and Microsoft are also exploring the potential of integrating gaze-typing~\cite{zhao2023gaze}. Additionally, Apple has announced that gaze-typing will be implemented in iOS 18 (beta)~\cite{apple_news_2024} to assist users with disabilities. As this technology becomes more prevalent, a broader range of devices will be susceptible to \system.

\section{Preliminary Attack Analysis}
\label{sec:attack}

\begin{figure*}[ht]
    \centering
    \includegraphics[width=\textwidth]{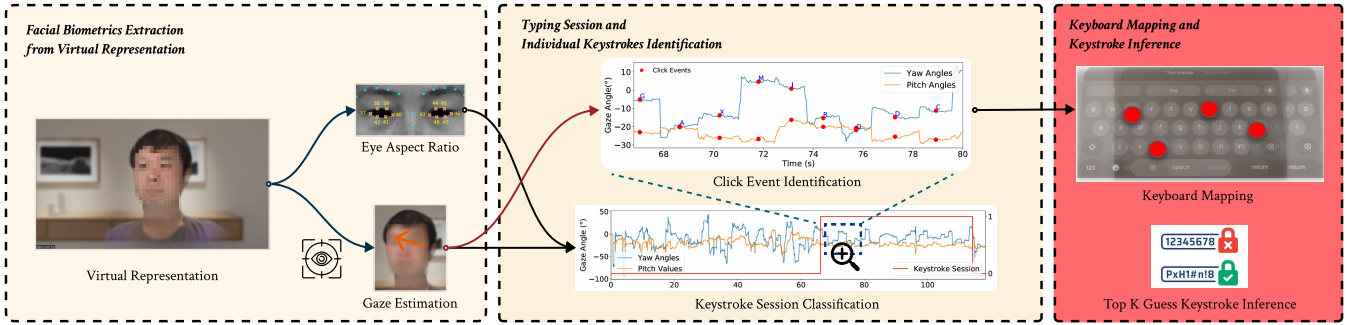}
    \caption{\textbf{\system} Attack Overview: The attacker starts by extracting
        facial biometrics from the victim's virtual representation, focusing on gaze
        estimation and eye aspect ratios (EARs). These biometrics are then used to
        distinguish typing sessions from other activities and to identify the timing
        of each keystroke. Finally, the attacker maps the gaze vectors to a virtual
        keyboard to infer the pressed keys.}
    \label{fig:overview}
\end{figure*}

This section evaluates the risk and explores various attack techniques,
emphasizing the realistic possibility of executing keystroke inference attacks. We have
developed the \system attack to tackle these challenges and include a
demonstration of its initial effectiveness. The attack process is detailed in
the following subsections. Section~\ref{sec:4.1} outlines our methods for
extracting biometrics such as eye aspect ratios (EARs) and gaze direction from
video frames. In Section~\ref{sec:4.2}, we discuss our technique for identifying
typing sessions using these metrics and describe our approach to determine the
timing of keystrokes. Section~\ref{sec:4.3} explains how we project gaze
directions onto a virtual keyboard plane to identify pressed keys, while
Section~\ref{sec:4.4} introduces a probability calculation technique for
inferring likely keystrokes. Finally, Section~\ref{sec:4.5} details a unique
mapping strategy for passcode inference.

In this section, we utilize the data that was collected on us and perform study
for experiment purpose. Specifically, we recorded demo videos while the authors
wearing and using Apple Vision Pro in a normal campus scenario. The videos
includes several scenarios such as browsing a webpage, typing in Slack,
attending Zoom meeting, unlock devices with passcodes, etc. We further randomly
choose a demo video with the duration of 120 seconds among them to analyze the
keystroke inference attack.

To begin with, we find successfully inferring
keystrokes based on very limited information of a video that only showing the
users' virtual face can be difficult. Specifically, we identify the following
three main challenges:

\begin{tcolorbox}[
        skin=widget,
        left=0mm,
        right=0mm,
        top=0mm,
        bottom=0mm,
        boxrule=0.5mm,
        colback=red!5,
        colframe=red!75!black!80,
        width=(\linewidth),
        segmentation style={solid, line width=0.35mm, left color=blue!15!yellow, right color=red!85!black, dashed},
        title=\hliwona{Main Challenges},
    ]

    \hliwona{Challenge 1: How to extract the essential biometrics?}

    \tcbline

    \hliwona{Challenge 2: When do users type?}

    \tcbline

    \hliwona{Challenge 3: Where is the virtual keyboard and which key is being pressed?}

\end{tcolorbox}

\subsection{Eye-related Biometrics Extraction}
\label{sec:4.1}

To address \textbf{Challenge 1}, we need to find suitable eye-related biometrics
besides gaze estimation. Neurologists already confirm that eye blink rate
decreases with increasing attentional demand~\cite{maffei2018spontaneous}. Given
typing in VR is typically an activity requiring high attention level, we expect
a lower eye blink rate during typing sessions. So we choose Eye Aspect
Ratio(EAR) as the biometric to measure eye blink rate. And thus our \system
attack relies on two biometrics, EAR and eye gaze estimation, as shown in
Figure~\ref{fig:overview}.

\

\noindent\textbf{Eye Aspect Ratio (EAR)} To quantify this, we use the 68-point
face landmark model to calculate. The EAR, calculated as the eye width, which is
the Euclidean distance between the eye corners, divided by the eye height, the
distance between the midpoints of the upper and lower eyelids, serves as a
metric for blink detection. A eye blink event is identified when this ratio
reach a local maximum as shown in Figure~\ref{fig:prelim_trace}.

\noindent\textbf{Gaze estimation} Our gaze estimation model employs a
pre-trained Resnet18 model~\cite{githubrepo}, which takes video frames with
human faces as input and output the yaw and pich angle values of eye gaze. The
model is trained on the ETH-XGaze dataset~\cite{zhang2020eth}. This dataset
features a wide range of head poses with yaw and pitch angles up to
$\pm$80$^{\circ}$,$\pm$80$^{\circ}$ and gaze yaw and pitch angles reaching
$\pm$120$^{\circ}$,$\pm$70$^{\circ}$ respectively. This range is broader
compared to other well-known datasets such as MPIIGaze~\cite{zhang2017mpiigaze}
or RT-GENE~\cite{fischer2018rt}, making it particularly suitable for our needs.
This is because in a virtual reality setting, the virtual camera's position
corresponds to the app's window, which can be located anywhere around the user,
not just directly in front. This flexibility ensures that our model accurately
captures gaze direction under various user orientations relative to the camera.
We also evaluate how robust the \system attack is regarding the position of the
Persona virtual camera in Section~\ref{sec:6.3}.

\subsection{Typing Activity Identification}
\label{sec:4.2}

We divide \textbf{Challenge 2} into two smaller problems: (1) how to identify
typing sessions, in which users type a string of characters consecutively, and
(2) how to identify the timings of individual keystrokes in these typing
sessions. To address these problems, we need to extract higher-level features
from the selected two biometrics, gaze estimation and EAR. Specifically, we
utilize an RNN for identifying typing sessions and design filters to extract
fixations from gaze estimation. Individual keystrokes are very likely happening
on these fixations lasting sufficient time.

\

\begin{figure}[htbp]
    \centering
    \includegraphics[width=\columnwidth]{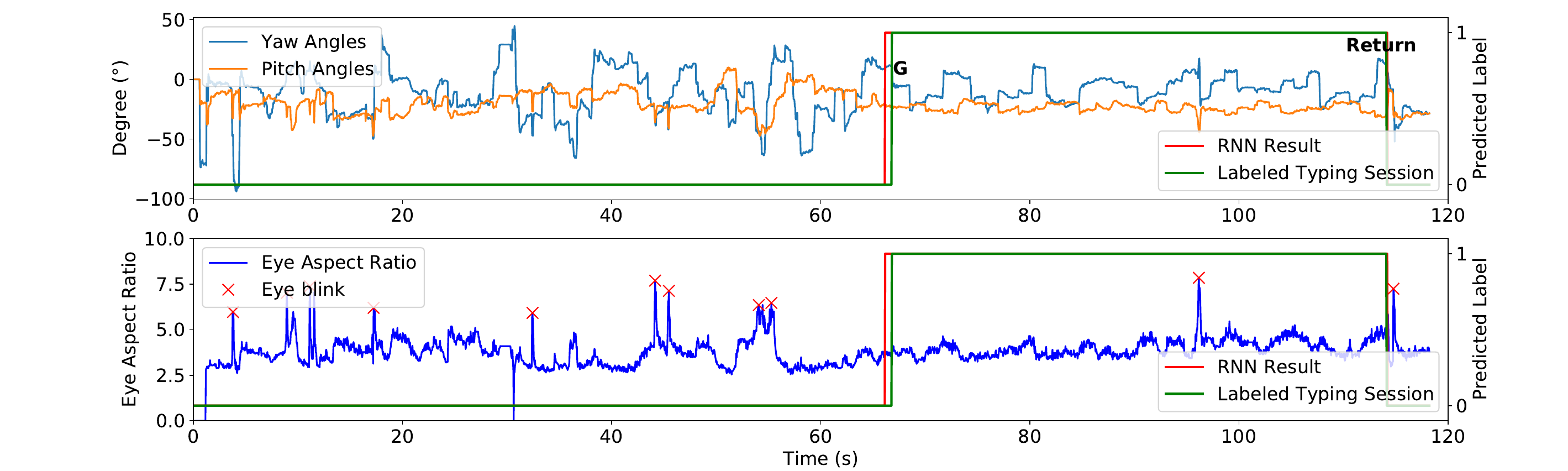}
    \caption{Eye gaze estimation and Eye Aspect Ratio of user a) browsing a webpage b) typing in iMessage, beginning with 'G' and ending with 'Return' c) sending the message and closing the apps}
    \label{fig:prelim_trace}
\end{figure}

\noindent\textbf{Identify Typing Sessions}  The gaze estimation depicted in
Figure~\ref{fig:prelim_trace} reveals a noteworthy pattern: during consecutive
keystrokes on a virtual keyboard, the direction of eye gaze tends to be more
concentrated and exhibits a periodic pattern, in contrast to the gaze patterns
during other activities, which appear more erratic and random. Further analysis
indicated that the frequency of eye blinking decreases during typing sessions
compared to other regular activities due to high attention demands. During
typing, eye blinks were significantly rarer. These observations prompted us to consider Recurrent Neural
Networks, RNNs, as a suitable method for distinguishing typing actions from
other activities. RNNs excel at recognizing patterns in sequential data, such as
the traces of eye gaze direction, making them exceptionally well-suited for
tasks where the context or sequence of inputs is
crucial~\cite{schuster1997bidirectional}.

\begin{figure}[htbp]
    \centering
    \includegraphics[width=\columnwidth]{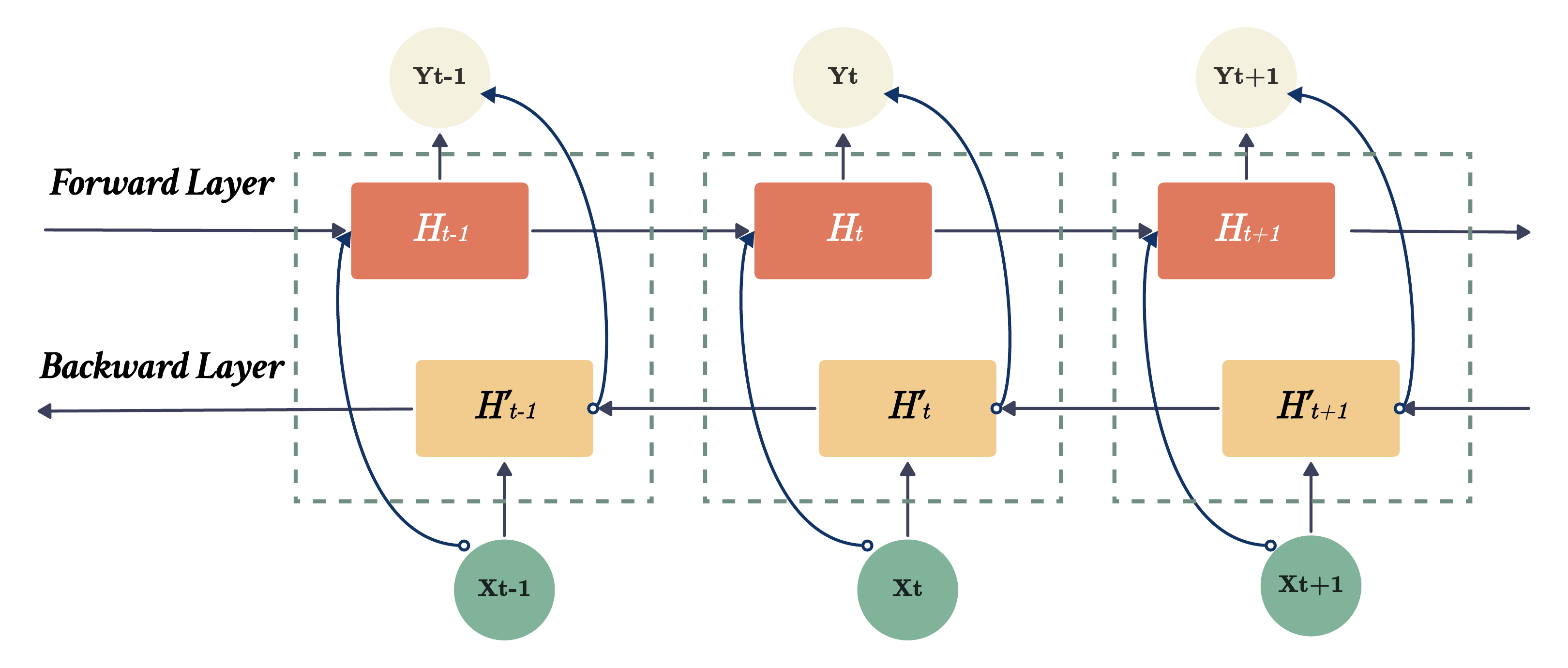}
    \caption{Bidirectional RNN Architecture for classifying typing sessions}
    \label{fig:rnn}
\end{figure}

We define a typing session as the period between the saccade of the first keystroke and the saccade of the last keystroke, as depicted in Figure~\ref{fig:prelim_trace}. We labeled the typing sessions manually and later use the labeled period in training and validation stages. We utilized a Bidirectional RNN model with its architecture shown in
Figure~\ref{fig:rnn}, specifically designed to classify typing events from other
VR usage scenarios. This model processes inputs representing the yaw and pitch
angles of eye gaze and eye aspect ratio. The model has a hidden layer size of
128, and differentiates between two classes: typing as label 1 or other
activities as label 0. The RNN is using cross entropy as the loss function and
the Adam optimizer to update its weights. This output will segment the raw gaze
trace into several clips labeled with the two classes. The model was trained
using 18 videos of the collected 23 videos in the comprehensive dataset. All the
typing sessions of the videos are labeled during the collection as described in
Section~\ref{sec:expe}. The accuracy of this model is detailed in
Section~\ref{sec:6.1}, and its application to the demonstration video is
illustrated in Figure~\ref{fig:prelim_trace}. In the demo attack, the user first
browses a webpage for a minute and then switches to iMessage from 0s to 66.8s.
Then the user inputs 35 letters from 66.8s to 114.1s, followed by sending the
message and closing the app. It can be observed that during the predicted typing
session, the yaw and pitch angles are within a range of
(40$^{\circ}$, 20$^{\circ}$) while in other activities the yaw and pitch surpass
(100$^{\circ}$, 50$^{\circ}$). There is also a periodic pattern of gaze staying
at each fixation for approximately 1s. Besides, the eye blink rate is much lower
when typing, occurring only 1 time in 40 seconds, whereas typically there is
about one blink every 7 seconds during other activities. These observations are
crucial features for identifying typing sessions.

\

\noindent\textbf{Identify Individual Keystrokes} 
After typing session identified, the next step is to recognize individual keystrokes. Clicks during gaze typing are confirmed by a snipping hand gesture.
While this gesture can be captured by the headset's external camera and displayed in the Persona avatar view, it is not always visible because the Persona view typically shows the user from the chest up, and users often rest their hands out of view. Therefore, we cannot rely on hand gestures to identify keystrokes. Instead, we use consistently visible features: saccades and fixations~\cite{blascheck2017visualization}. During gaze typing, users' gazes shift between keys and fixate on the key to be clicked, resulting in saccades followed by fixations. We developed an algorithm to detect these patterns in the gaze traces.

\begin{figure}[htbp]
    \centering
    \includegraphics[width=\columnwidth]{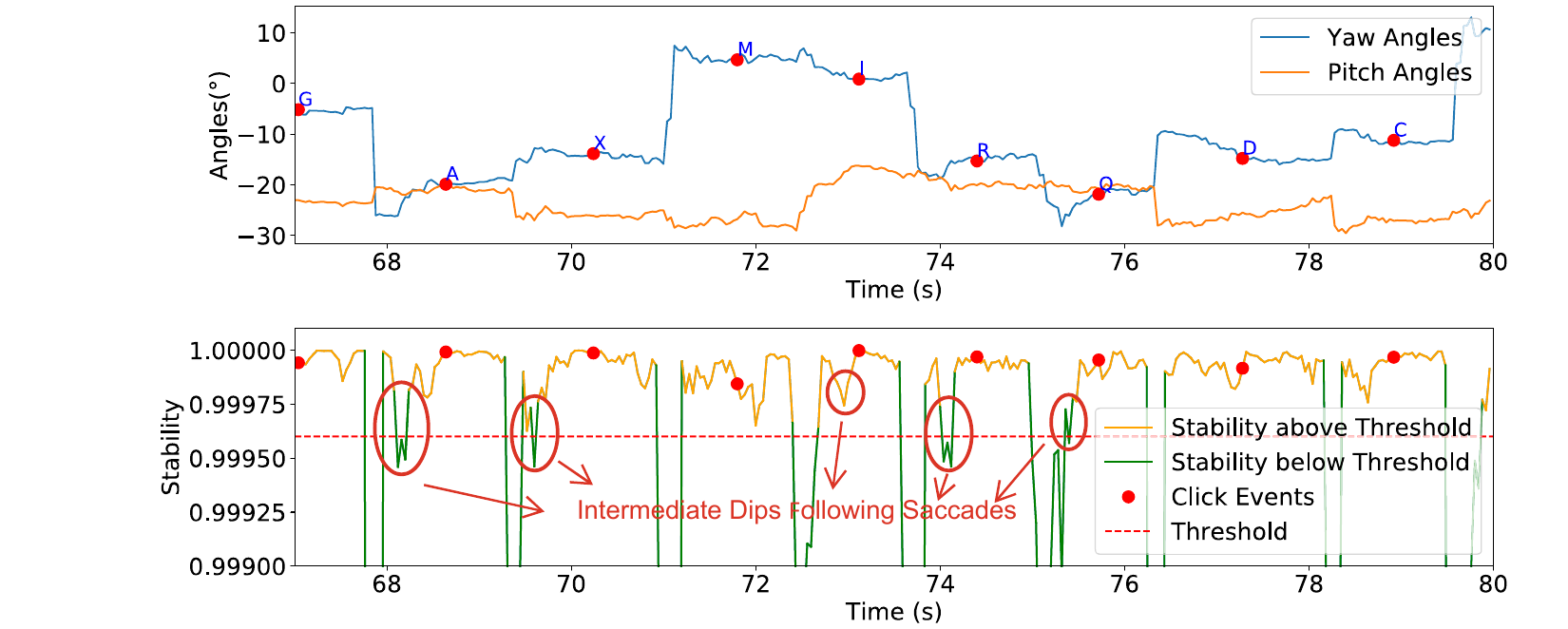}
    \caption{Identify individual keystrokes in the keystroke session. The red dashed line is the $S_{T}$ to distinguish saccades (green traces) and fixations(yellow traces).}
    \label{fig:keystroke_detect}
\end{figure}

First, we calculate gaze stability $S$, the average cosine similarity between gaze directions within a time window of length $N$, as shown in Equation~\ref{eq:cosSimi}. In the equation, $\alpha_{i,j}$ denotes the angle between two gaze directions ($\phi_i$,$\theta_i$) and ($\phi_j$,$\theta_j$). As illustrated in Figure~\ref{fig:keystroke_detect}, saccades induce lower stability and fixations induce higher stability.

\begin{equation}
    \label{eq:cosSimi}
    \begin{split}
         & {S}(n) = \frac{1}{N^2} \sum_{i=n}^{n+N-1} \sum_{j=n}^{n+N-1} \cos(\alpha_{i,j})                          \\
         & \cos(\alpha_{i,j}) = \cos(\phi_i) \cos(\phi_j) \cos(\theta_i - \theta_j) + \sin(\phi_i) \sin(\phi_j)
    \end{split}
\end{equation}

Saccades and fixations can be distinguished by analyzing dips in the stability trace. Saccades cause large dips, while noise during fixations results in small dips. To differentiate between them, we need a threshold $S_T$ that captures most saccade-induced dips while excluding noise-induced ones. We observed that saccades are often followed by minor gaze adjustments that create intermediate dips, which are smaller than saccade-induced dips but larger than noise-induced dips. An effective strategy is to select $S_T$ within the range of these intermediate dips, as it includes most saccade-induced dips while rejecting noise-induced ones. Although this threshold captures some intermediate dips from post-saccade adjustments, they can be filtered out based on their timing, as they always closely follow deeper saccade-induced dips.

\begin{figure}[htbp]
    \centering
    \includegraphics[width=\linewidth]{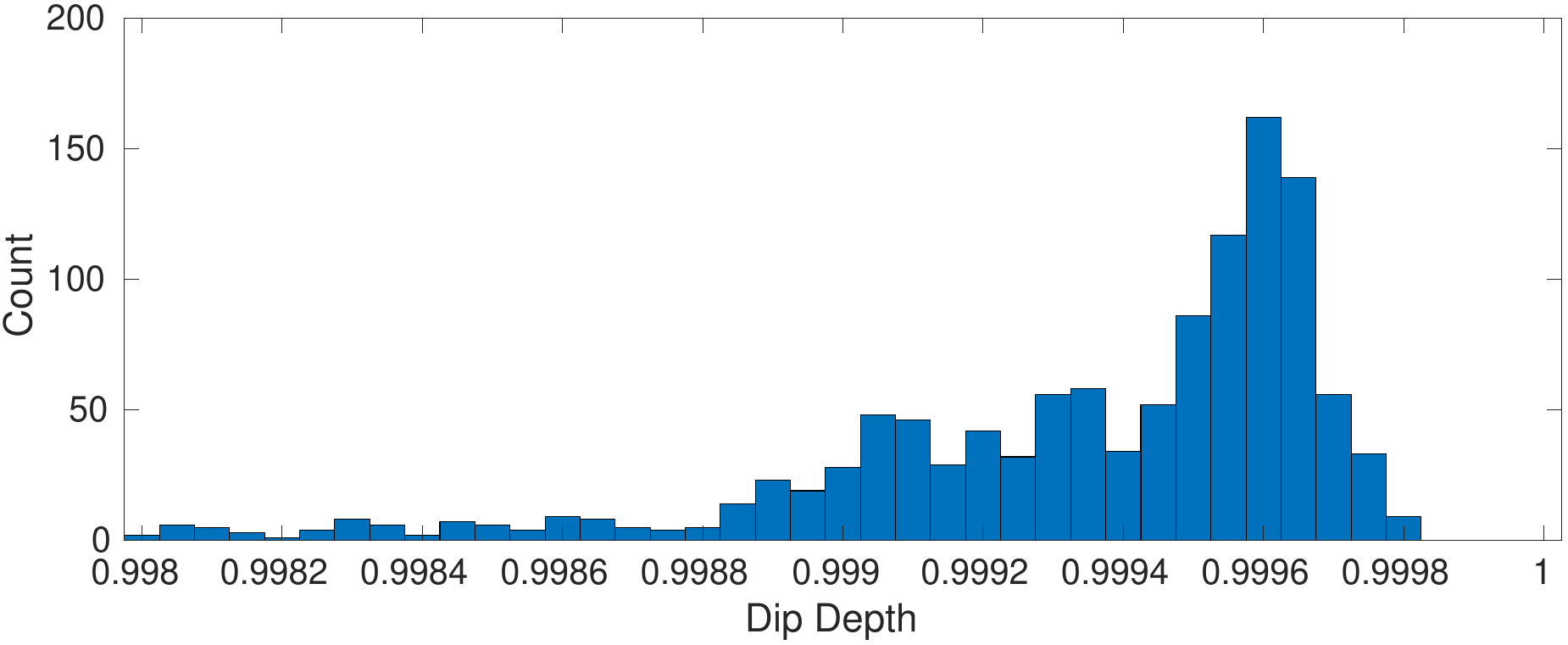}
    \caption{Averaged Optimal Threshold Distribution}
    \label{fig:threshold_distribution}
\end{figure}

To estimate the optimal $S_T$, we use function \texttt{findpeaks($-S$)} in MATLAB, setting the \texttt{MinPeakHeight} parameter to \texttt{median$(-S)$} to identify significant dips in the trace. We then analyze each trace to find all intermediate dips, defined as those occurring within 500 ms of a deeper dip, as shown in Figure~\ref{fig:keystroke_detect}. 
We identify all dip depth ranges that consist entirely of intermediate dips and select the range with the highest number of intermediate dips. 
This range is considered the optimal threshold range, and we record the depths of all dips within it. 
We repeat this process for all data, recording dip depths in the optimal threshold range for all traces. 
We then plot the distribution of all recorded dip depths in Figure~\ref{fig:threshold_distribution}. 
The results show that selecting 0.9996 as $S_T$ is most optimal since it aligns with the peak of the overall intermediate dip depth distribution.

After determining $S_T$, we apply it to the gaze traces to select dips deeper than $S_T$ as saccade candidates. 
We then remove intermediate dips that occur within 500 ms after deeper dips, as these are typically caused by minor adjustments.
Finally, the intervals between the preserved saccades are identified as fixations, each corresponding to a single keystroke click.
We show part of the identified clicks in the demo attack in
Figure~\ref{fig:keystroke_detect}. As illustrated in the figure, blue labeled letters
indicate the correctly identified clicks. The overall precision and recall
of the click identification on the collected dataset are presented in Section~\ref{sec:6.1}.

In our demo attack, there were 35 actual keystrokes. The attack model identified
36 keystrokes in the trace, comprising 35 true positives and 1 false positive,
with no false negatives. This resulted in a precision of 97.2$\%$ and a recall
of 100$\%$ for the demo attack.

\subsection{Adaptive Virtual Keyboard Layout Mapping}\label{sec:4.3} 
\begin{figure}[htbp]
    \centering
    \includegraphics[width=\columnwidth]{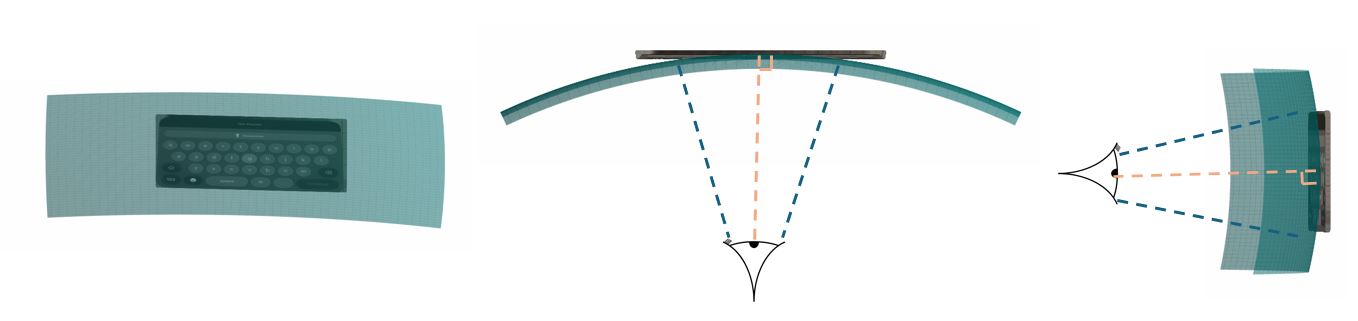}
    \caption{Virtual Keyboard is on a sphere facing the center at the user's eyes. The normal vector of the keyboard points to the midpoint of the user's eyes. }
    \label{fig:keyboardDemo}
\end{figure}

To
accurately map gaze points to specific keys during individual keystrokes, it is
essential to precisely determine the location of the virtual keyboard in virtual
space. Typically, the keyboard is a rectangular area on a plane, with a line
connecting its center to the user acting as the plane's normal vector, as shown
in Figure~\ref{fig:keyboardDemo}. Users are free to adjust the
keyboard's size, direction, and distance within the virtual space, causing its
position to vary widely on a sphere. While these specific adjustments are not
directly observable to an attacker, this section explains how we can utilize
eye-movement statistics to accurately estimate the keyboard's location.

\noindent\textbf{Keyboard Plane Estimation} We first estimates the plane where
the virtual keyboard sits in the virtual space. Because the line connecting the
user to the keyboard center is perpendicular to this plane, the average gaze
direction can estimate the plane's normal vector. Subsequently, the gaze
directions are rotated into a new coordinate system where the estimated normal
vector of the keyboard plane aligns with the $x$-axis. The rotated gaze
directions described by yaw angle $\theta$ and pitch angle $\phi$ are then
transformed into gaze directions represented by the unit vector $\mathbf{u}$ in
the XYZ coordinate using Equation~\ref{eq:polar2xyz}.
\begin{equation}\label{eq:polar2xyz}
    \hat{\mathbf{u}} =
    \begin{bmatrix} u_x \\ u_y \\ u_z \end{bmatrix}
    =
    \begin{bmatrix}
        \cos(\theta) \cos(\phi) \\
        \sin(\theta) \cos(\phi) \\
        \sin(\phi)
    \end{bmatrix}
\end{equation}

The keyboard plane is parallel to the $y$-$z$ plane. The gaze direction can be
projected to the gaze points on the keyboard plane using
equation~\ref{eq:gazepoints}, with $p_x$ being the distance between the keyboard
and the user. $\mathbf{p}$ represents the gaze point coordinates on the keyboard
plane in the $y$-$z$ space.
\begin{equation}\label{eq:gazepoints}
    \mathbf{p}
    = \begin{bmatrix} p_y \\ p_z \end{bmatrix}
    = \begin{bmatrix} u_y \\ u_z \end{bmatrix} \times \frac{p_x}{u_x}
\end{equation}

\ 

\noindent\textbf{Keyboard Area Estimation} After determining the keyboard plane,
we must also identify the specific area within that plane where the keyboard is
located. The keyboard typically has a much wider horizontal span than vertical,
leading to a predominance of horizontal gaze direction movements during typing.
Consequently, the orientation of the keyboard can be estimated by analyzing the
distribution of gaze points. Using Principal Component Analysis (PCA), we can
identify the principal directions of variance in the dataset, which corresponds
to the horizontal direction of the keyboard.

This involves computing the eigenvectors and eigenvalues of the dataset's
covariance matrix. First, we compute the covariance matrix \(\mathbf{C}\) of the
dataset using Equation~\ref{eq:covariance}
\begin{equation}\label{eq:covariance}
    C = \frac{1}{n-1} (\mathbf{P} - \overline{\mathbf{P}})^T (\mathbf{P} - \overline{\mathbf{P}})
\end{equation}
where \(\mathbf{P}\) represents the matrix of all data points \(\mathbf{p}_i\) and \(\overline{\mathbf{P}}\) is the mean vector of these points.
Eigenvectors \(\mathbf{v}_i\) and their corresponding eigenvalues \(\lambda_i\) are computed using Equation~\ref{eq:eigenvec}
\begin{equation}\label{eq:eigenvec}
    C\mathbf{v}_i = \lambda_i\mathbf{v}_i
\end{equation}
The eigenvector with the highest eigenvalue indicates the direction of the greatest variance, i.e., the horizontal direction.
To reorient the original dataset into principal components, the data points are
projected onto these eigenvectors using Equation~\ref{eq:project_to_eigenvec}
\begin{equation}\label{eq:project_to_eigenvec}
    \mathbf{p}' = \mathbf{V}^T \mathbf{p}
\end{equation}
where \(\mathbf{V}\) comprises the columns formed by the eigenvectors
\(\mathbf{v}_i\) of \(\mathbf{C}\).
This transforms the gaze points to $\mathbf{p}'$
so that the transformed axes align with the horizontal and vertical directions
of the keyboard.

\begin{figure}[htbp]
    \centering
    \includegraphics[width=\columnwidth]{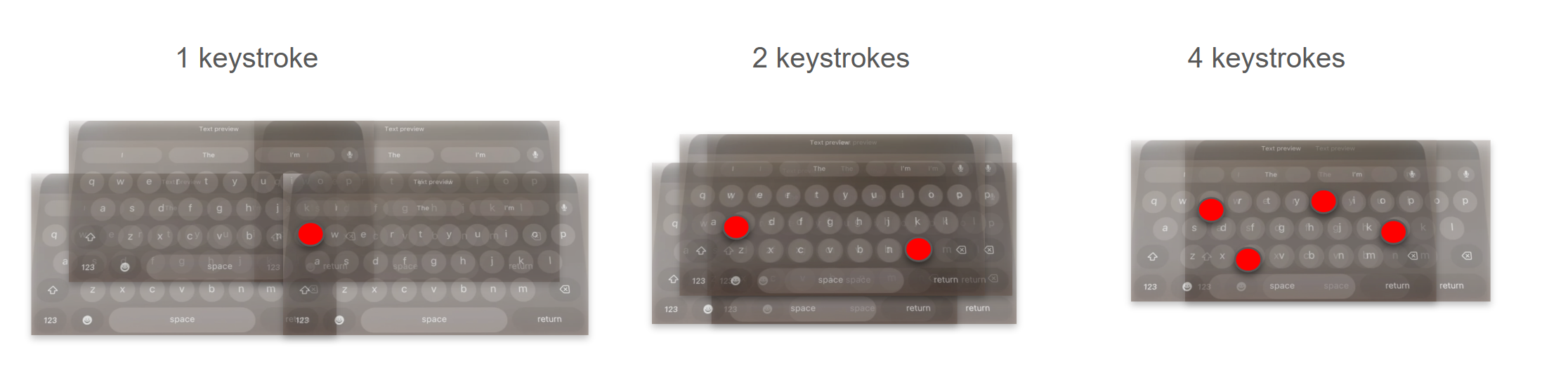}
    \caption{Virtual Keyboard possible locations decrease with more edge-located keystroke angle information.} %
    \label{fig:keyboard_locate}
\end{figure}

After identifying the keyboard orientation, the keyboard boundary can be
estimated using the distributions of the gaze points. We hypothesize that the
gaze points extend linearly across the maximum possible area of the keyboard,
spanning horizontally from the letter Q to the letter P, and vertically from the
letter Y to the Space key. Initially, this assumption may seem questionable, as
users might typically focus on a limited section of the keyboard. However, upon
evaluating the four typing scenarios, this approach is justified by very high
key coverage on the edge of the keyboard. For instance, in message inference
scenarios, the Space key is frequently used. The message is concluded with a
press of the return key. Further validation comes from analyzing two Wikipedia
pages. Out of 676 sentences
longer than 15 characters, only 8 lacked the leftmost letters ``\texttt{q}'',
``\texttt{a}'', or ``\texttt{z}''. In password inference scenarios, presses of
the return, numberspace, and number keys must be included as modern passwords
require a combination of letters and numbers. For URL inference, the
``\texttt{w}'' letter and ``\texttt{.}'' and the return key are usually
included. All these keys are edge-located keys that guarantee the typing area is
the whole keyboard instead of a limited region. As shown in
Figure~\ref{fig:keyboard_locate}, with more than 4 edge-located keystrokes
spanning both horizontally and vertically, we can locate the keyboard within the
error of the distance of two adjacent keys.

\subsection{Top K Keystroke Candidate Accuracy}\label{sec:4.4}
\begin{figure}[htbp]
    \centering
    \subfloat[]{\includegraphics[width=0.45\columnwidth]{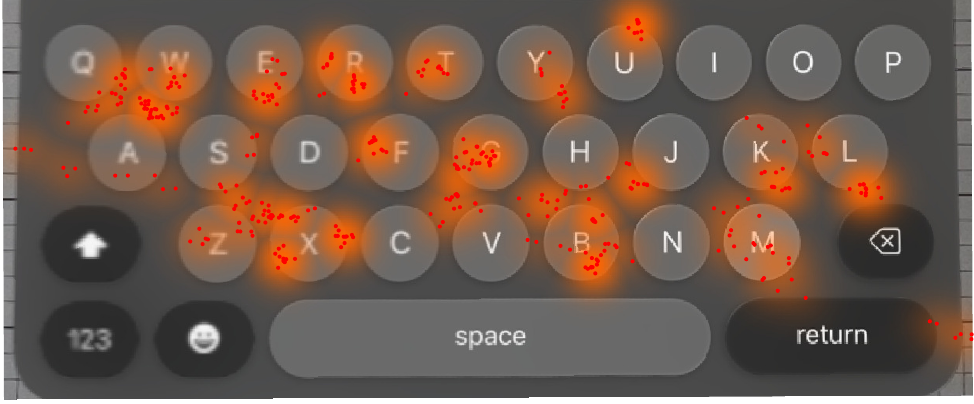}}

    \subfloat[Predicted]{\includegraphics[width=0.45\columnwidth]{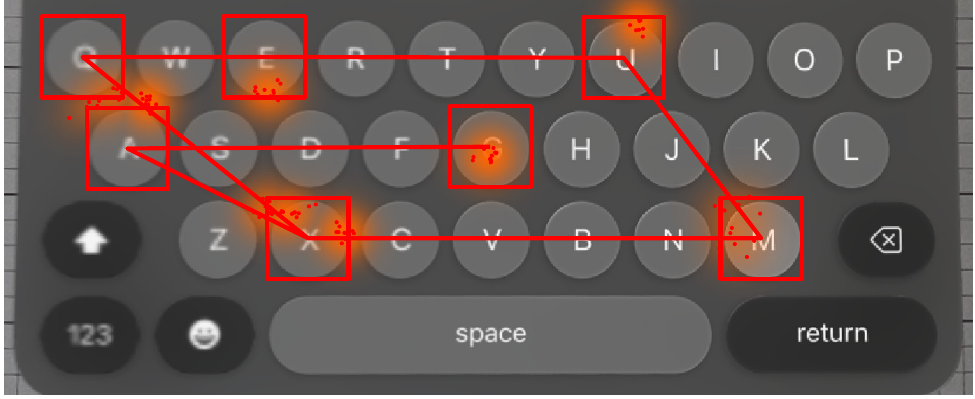}}
    \subfloat[True]{\includegraphics[width=0.45\columnwidth]{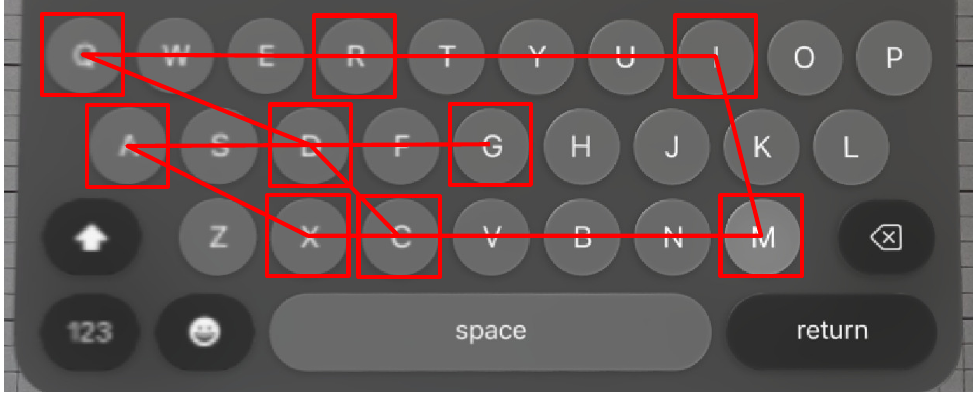}}

    \caption{Mapped gaze directions for keystroke inference of the demo attack (a) Adaptive virtual keyboard mapping (b) Predicted first guess keystrokes from 67-80s (c)Actual keystrokes (True Labels) from 67-80s}
    \label{fig:keystroke_mapping}
\end{figure}

After locating the keyboard, the gaze points of the individual keystrokes
are mapped to the keyboard layout as shown in
Figure~\ref{fig:keystroke_mapping}. We assume a 2D Gaussian probability
distribution centered at the gaze direction for each frame in a fixation of one
individual keystroke, the distribution has a 2$\sigma$ value of the radius of
one key. The probability of each key is averaged across every frame in each
candidate keystroke period as depicted in Equation~\ref{eq:probability}. In the
equation, $P(C_{k})$ denotes the probability of click event $C$ inferred as key
$k$, $N_{C}$ denotes frame numbers in the fixation duration of one individual
keystroke, $\mathcal{N}_{i}$ denotes the 2D Gaussian distribution of the gaze in
the $i$th frame, $S_{k}$ denotes the 2D region of key $k$. $p_{y_{i}},
    p_{z_{i}}$ are the horizontal and vertical coordinates of the gaze direction
mapping in the $i$th frame, $\sigma_{y_{i}},\sigma_{z_{i}}$ are horizontal and
vertical standard deviations, which are fixed values of half the radius of a
letter key.

\begin{equation}
    \label{eq:probability}
    P(C_{k}) = \sum_{i}^{i\in N_{C}} \frac{\iint_{S_{k}} \mathcal{N}_{i}((p_{y_{i}}, p_{z_{i}})  , (\sigma_{y_{i}},\sigma_{z_{i}}))}{N_{C}}
\end{equation}

\begin{figure}[htbp]
    \centering
    \includegraphics[width=\columnwidth]{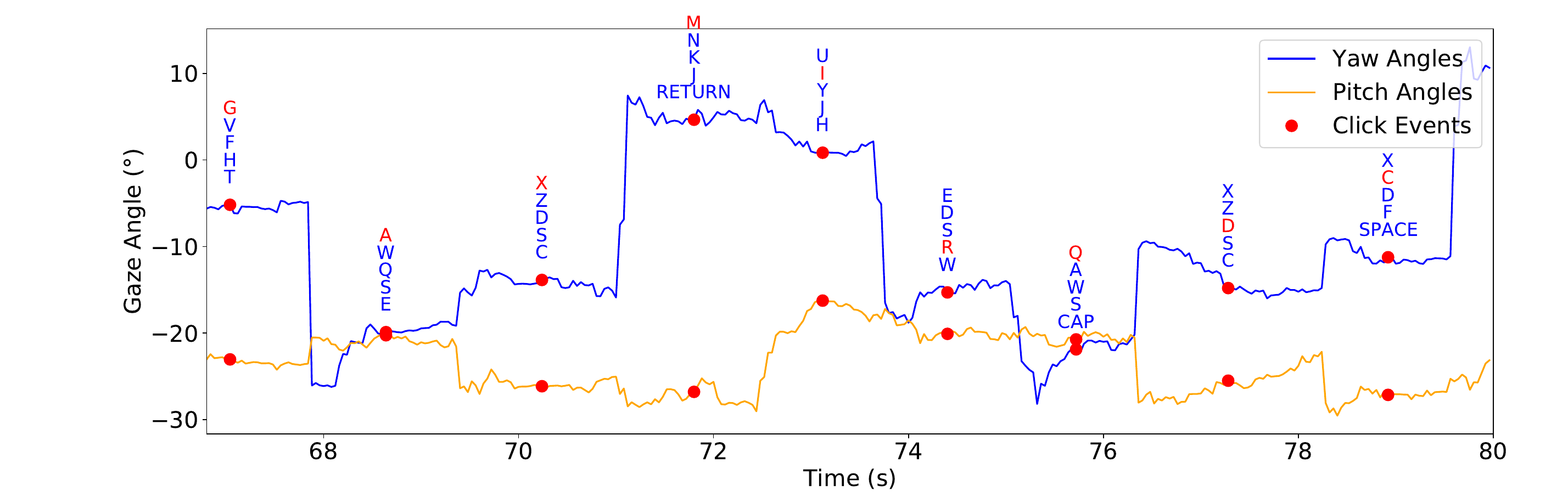}
    \caption{Top 5 guesses of keystroke letters listed from top to bottom of 67-80s in the keystroke session in the demo attack, red letters represent the actual keystroke labels}
    \label{fig:TopKdemo}
\end{figure}

We then look at the top K keystroke inference guesses that have the top K
highest probability, and infer the keystroke based on these candidates as shown
in Figure~\ref{fig:TopKdemo}. The red letter in blue 5 letters is the correct
label of the keystroke. In this demo attack, When K is set to be 5, the
character inference accuracy of the attack reaches 100$\%$.

\subsection{Passcode Inference}
\label{sec:4.5}

The passcode inference attack is executed by identifying the key patterns that
best match the observed gaze point patterns. This involves calculating a
translation vector $\mathbf{t}$ that minimizes the distances between the gaze
points and their nearest corresponding keys. The process is mathematically
represented in Equation~\ref{eq:mindistance}, where $\mathbf{p_i}$ represents
the coordinates of the $i$th gaze point and $\mathbf{k_j}$ represents the
coordinates of the $j$th key on the keyboard.
\begin{equation}\label{eq:mindistance}
    \mathbf{t} = \min_\mathbf{t} \sum_{i=1}^N \min_{j} \| (\mathbf{p_i} + \mathbf{t}) - \mathbf{k_j} \|^2
\end{equation}

After applying the calculated
translation vector to the gaze points, the probability of each key being typed
can be calculated using Equation~\ref{eq:probability}, and the pressed key can
be inferred.

\begin{figure}[htbp]
    \centering
    \subfloat[Predicted]{\includegraphics[width=0.24\columnwidth]{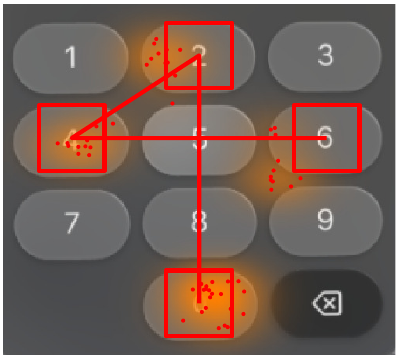}}
    \subfloat[True]{\includegraphics[width=0.24\columnwidth]{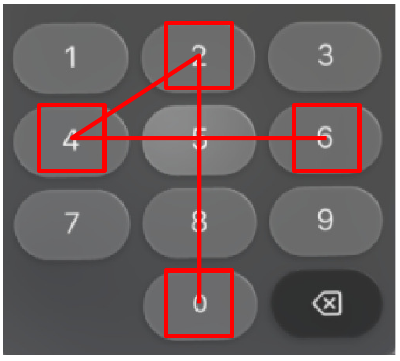}}
    \subfloat[Predicted]{\includegraphics[width=0.24\columnwidth]{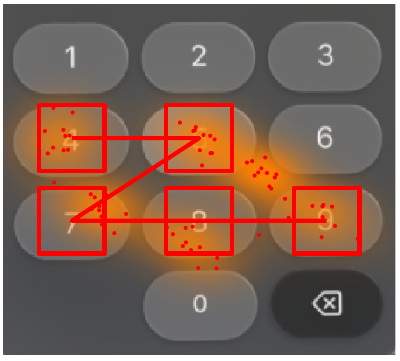}}
    \subfloat[True]{\includegraphics[width=0.24\columnwidth]{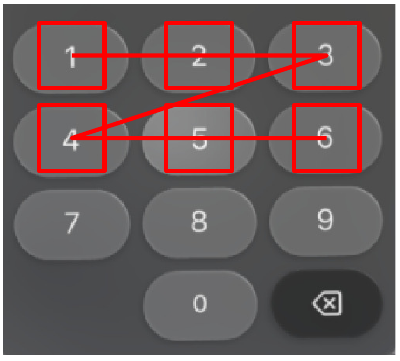}}

    \caption{Mapped gaze points and the first guess inferred passcode for keystroke inference on the PIN keyboard, (a) can be correctly inferred within one guess to match (b), (c) can be correctly inferred within four guesses to match (d).}
    \label{fig:passcodemapping}
\end{figure}

Two attack examples are shown in Figure~\ref{fig:passcodemapping}. A potential issue arises when multiple passcodes share the same pattern. If the pattern is unique, as in Figure~\ref{fig:passcodemapping} (a) and (b), the correct passcode might be guessed on the first attempt. However, if the same pattern corresponds to multiple passcodes, additional guesses may be necessary, as shown in Figure~\ref{fig:passcodemapping} (c) and (d). Nonetheless, this issue only slightly increases the passcode search space.

\section{VR Gaze-typing Data Collection}
\label{sec:expe}

\subsection{Comprehensive Dataset}
We began data collection only after receiving approval from the Institutional
Review Board (IRB) to further validate our attack designs in previous sections. Our
study included 23 participants representing diverse racial, gender, and cultural
backgrounds. Prior to data collection, participants were provided with an
information form detailing the tasks involved, potential risks, and measures
taken to ensure data confidentiality, as per IRB guidelines. Participants were
required to sign this form before proceeding.

During the initial phase, all participants underwent eye-tracking feature
recalibration on the Apple Vision Pro device and updated their Persona view.
Subsequently, they joined a Zoom meeting from their VR headset, sharing their
Persona view in Zoom to commence recording. The duration of the recorded
meetings averaged 16 minutes and 10 seconds per participant, constituting the
raw data.

Participants were instructed to complete seven tasks during the recorded
meeting, presented in a randomized order to compose a comprehensive dataset: (1) Open a web browser, navigate to \texttt{www.google.com}, enter the Gmail address   and password to sign in to a
      test Google account, perform a Google search of arbitrarily chosen queries, and browse the search results. Then, navigate to
      another commonly used website. (2) Watch a 3-4 minute video on Apple TV. (3) Open Slack and send several messages. (4) Engage in conversation with Zoom participants who have their cameras
          on. (5) Engage in conversation with individuals in the physical world,
          situated close to the participants. (6) Play Super Fruit Ninja. (7) Change the passcode.

For all text entry tasks, participants are instructed to indicate the beginning
and end of their typing sessions, which serves as a reference for labeling the
corresponding video segments during the typing session classification phase. 
To label the typing sessions, we manually check the gaze estimation traces around the reference timings of all participants to find the saccade pairs of the first and last keystrokes of each typing session. The time intervals between these saccade pairs are labeled as typing sessions (Label 1). All other time intervals are labeled as other activities (Label 0).
Within each typing session, we further label the keystrokes with the correct keys being clicked. This detailed labeling is used to evaluate the accuracy of keystroke inference. In the comprehensive dataset, the typing session duration averaged 2 minutes and 25 seconds per participant.

During text entry tasks in the web browser, participants navigate to
``\texttt{www.google.com}'' and optionally another commonly used website of their choice.
They use a Gmail account with the email address to be either
``participantvisionpro@gmail.com'' or ``visionproparticipant@gmail.com''.
Participants search for either a predetermined phrase (``\texttt{VR headset}'')
or a phrase of their choosing, which we subsequently verify through Google
search history as the correct label.

For message inference, participants are instructed to send a message to another
Slack account. This includes both a predetermined message (``\texttt{A quick
    brown fox jumps over the lazy dog}'') and a message of their own composition.
These messages sent to our Slack account serve as labels for correct keystrokes.

To facilitate password inference, participants select a random password in
adherence to Google account password requirements: (1) a minimum of 8
characters, (2) a combination of letters, numbers, and/or symbols, (3) no
leading or trailing blank spaces, and (4) not easily guessable (e.g.,
``\texttt{password123}''). The chosen password is then temporarily set as the
password for a test Google account. During the data collection process,
participants use this password to sign in, allowing us to capture their typing
behavior accurately. Regarding Passcode (PIN) inference, participants select a
6-digit PIN code prior to the experiment, enabling us to identify the correct
keystrokes during data analysis. To change the Passcode, each participant is required to input the original Passcode once and input the new Passcode twice to confirm the change. 3 participants did an extra round of Passcode input due to misclicks.
All video recordings are conducted in five distinct indoor scenarios over a
period exceeding four weeks.

In the dataset, 23 participants input 24 unique PINs, totaling 72 6-digit PIN entries. For password inference, there are 23 different password strings with an average length of 12.43 characters. In message inference, aside from the fixed sentence, the self-chosen sentences have an average length of 20.52 characters. The average length of the 40 collected email/URL strings is 19.8 characters. Participants are allowed to use ``\texttt{BACKSPACE}'' to correct the typos made in the dataset. There are 84 ``\texttt{BACKSPACE}'' keystrokes in the dataset.

\noindent\textbf{IRB approval and Responsible Disclosure} Our data collection
process has been approved by the university's Institutional Review Board (IRB).
We have reported these vulnerabilities to Apple. Apple Security Team has reproduced \system using our artifacts and assigned CVE-2024-40865. Patches are deployed on visionOS 1.3.

\subsection{Sentence Typing Dataset}
We collect an additional dataset specifically for sentence typing inference evaluations.
This dataset involves 15 participants, 7 of whom are not included in the initial group of 23, bringing the total number of participants to 30. 
The experimental design builds upon the sentence typing experiment used in Typose~\cite{goingMotions}, but with significant improvements.
Typose required participants to input 5-word ``sentences'' synthesized by random permutations from a pool of 60 selected words.
In contrast, our experiment asks each participant to type 15 different 7-word real sentences generated by ChatGPT. A total of 225 unique sentences are created, using words from the 10,000 most frequent English words found in the OpenSubtitles Dataset GitHub Repository~\cite{wictionary}, without any other restrictions on word selection.

\begin{figure}[htbp]
    \centering
    \includegraphics[width=\linewidth]{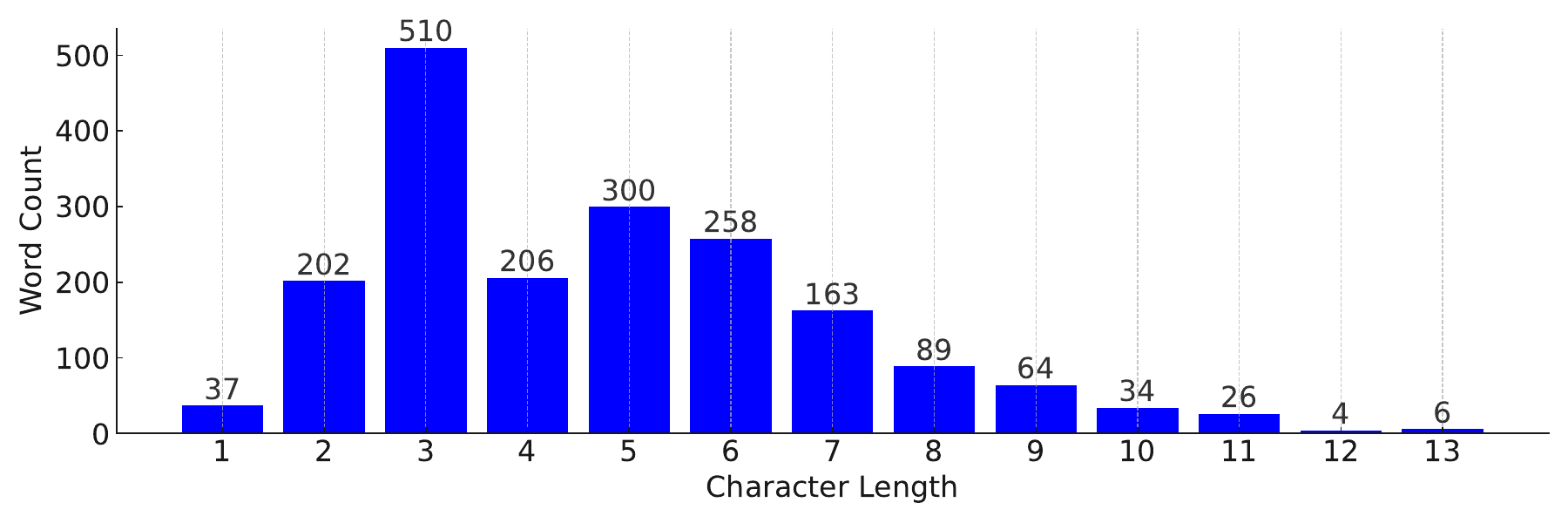}
    \caption{Word Counts of different word length}
    \label{fig:wordlength}
\end{figure}

In this experiment, video recordings averaging 21 minutes and 30 seconds in length are captured for each participant, with the typing sessions averaging 13 minutes and 15 seconds in length.
The dataset consists of 10329 keystrokes, including 1351 ``\texttt{SPACE}'' keys, 359 ``\texttt{BACKSPACE}'' keys, and 8619 English letters.
In total, it contains 1,575 words, of which 594 are unique.
Figure~\ref{fig:wordlength} shows the distribution of word lengths in the typed sentences, which range from 1 to 13 characters. 
Given that only 14 out of 10,000 words from the OpenSubtitles Dataset exceed 13 characters, our dataset is highly representative of the English vocabulary.

Compared to Typose~\cite{goingMotions}, our dataset better represents real-world attack scenarios in the following ways:
\\
\noindent\textbf{More unique words}: Our dataset contains 594 unique words, significantly more than the 60 used in Typose.
\\
\noindent\textbf{No word length constraints}: Our dataset includes words ranging from 1 to 13 characters in length, whereas Typose uses fixed-length words of 2 or 6 characters. 
\\
\noindent\textbf{More realistic sentences}: We use meaningful, grammatically correct sentences to better reflect real-world attack scenarios compared to the simple random word permutations used in Typose.
\\
\noindent\textbf{Allow typo corrections}: We allow participants to make and correct typos using the ``\texttt{BACKSPACE}'' key to reflect real-world typing behavior, whereas in Typose, words with typographical errors are considered unclassifiable.

\section{Evaluation}
\label{sec:eval}

The evaluation consists of two stages. The first stage assesses the system's
ability to detect when typing activity occurs—referred to as typing sessions and
individual keystroke identification, which addresses \textbf{Challenge 2}. The
second stage evaluates the accuracy of mapping the detected gaze to specific
keys on a keyboard, thus determining which keys were pressed—termed keystroke
inference accuracy addressing \textbf{Challenge 3}.

\subsection{Typing Activity Identification}
\label{sec:6.1}

We use precision, recall, and accuracy in Equation~\ref{eq:click_precision} as
metrics for typing session and individual keystroke identification results.
\begin{equation}
    \label{eq:click_precision}
    \begin{split}
         & Preceision = \frac{TP}{TP + FP} \text{ , }  Recall = \frac{TP}{TP + FN} \\
         & Accuracy = \frac{TP + TN}{TP + FP + TN + FN}
    \end{split}
\end{equation}

\

\begin{figure}[htbp]
    \centering
    \includegraphics[width=\linewidth]{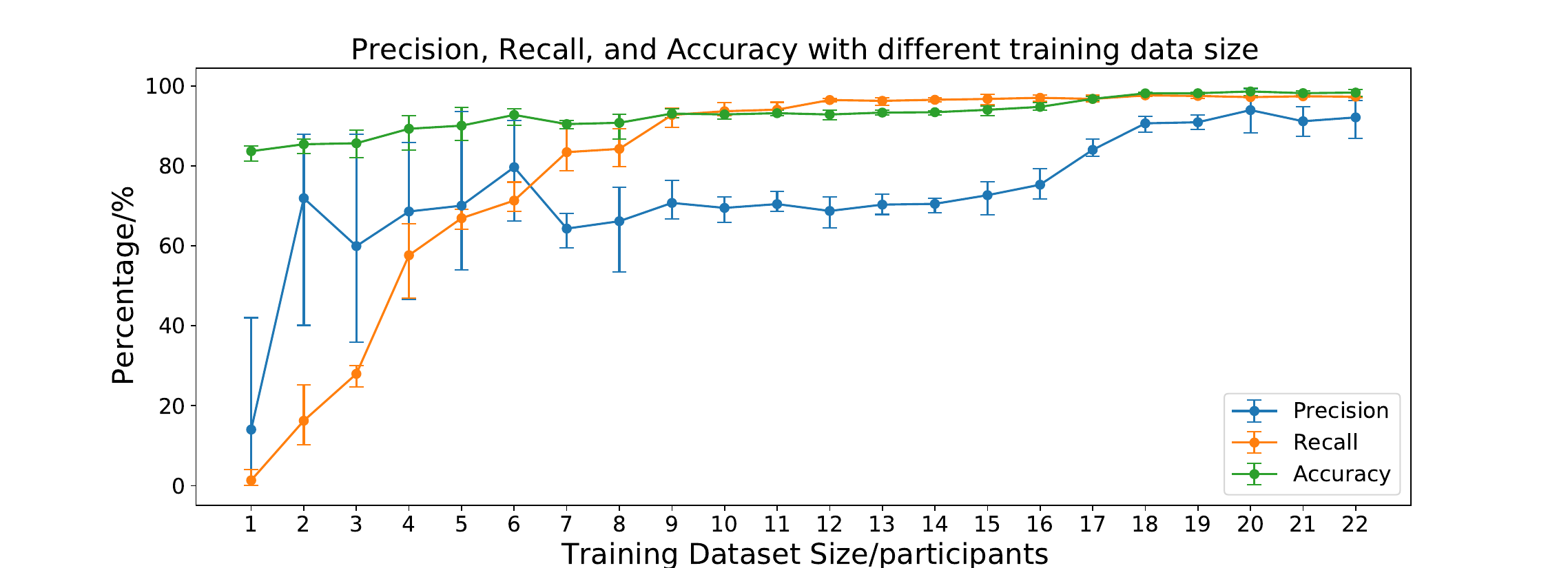}
    \caption{Accuracy, Precision, Recall vs. Participants as training dataset}
    \label{fig:trainingDataSize}
\end{figure}

\noindent\textbf{Typing Session identification results} 
We assume the attacker can first collect labeled data from non-victims and train an RNN
model for typing session identification in the attack. 
We use the data
with labels of typing sessions from some participants in the comprehensive dataset for training the RNN model and
validate it on both the remaining participants' data in the comprehensive dataset and also all data from sentence typing dataset. And we iterate through all the 23
participants and train and validate 23 rounds, following the K-fold
cross-validation style. Each round is trained with a batch number of 64 and
an epoch number of 100. We change the participants' number in the training dataset from 1 to 22 and plot the average, upper, and lower bound of precision, recall, and accuracy in Figure~\ref{fig:trainingDataSize}. The accuracy, precision, and recall will rise and stabilize around 98.1$\%$, 90.5$\%$, and 97.2$\%$ when the training dataset size exceeds 18 participants. This high accuracy result validates our theory on
combining both gaze vectors and eye aspect ratio to predict gaze typing sessions
in VR.

\noindent\textbf{Individual Keystrokes Identification Results}  
We evaluate the
precision and recall of the individual keystrokes identification method. We have
12839 actual clicks in identified typing sessions, 12428 of which were successfully identified as true
positives, \system also infers 2039 false positives where there are no clicks but
identified as a click. Overall the average precision rate and recall rate are 85.9\% and 96.8\%. There is a trade-off between precision and recall rate
because we use a threshold of gaze stability to identify keystrokes. A less
sensitive threshold value can lead to a lower recall rate and higher precision
rate and thus be beneficial in certain attack scenarios.

\subsection{Keystroke Inference Accuracy}

\begin{figure}[htbp]
    \centering
    \subfloat[K=1 (QWERTY)]
    {\includegraphics[width=0.45\columnwidth]{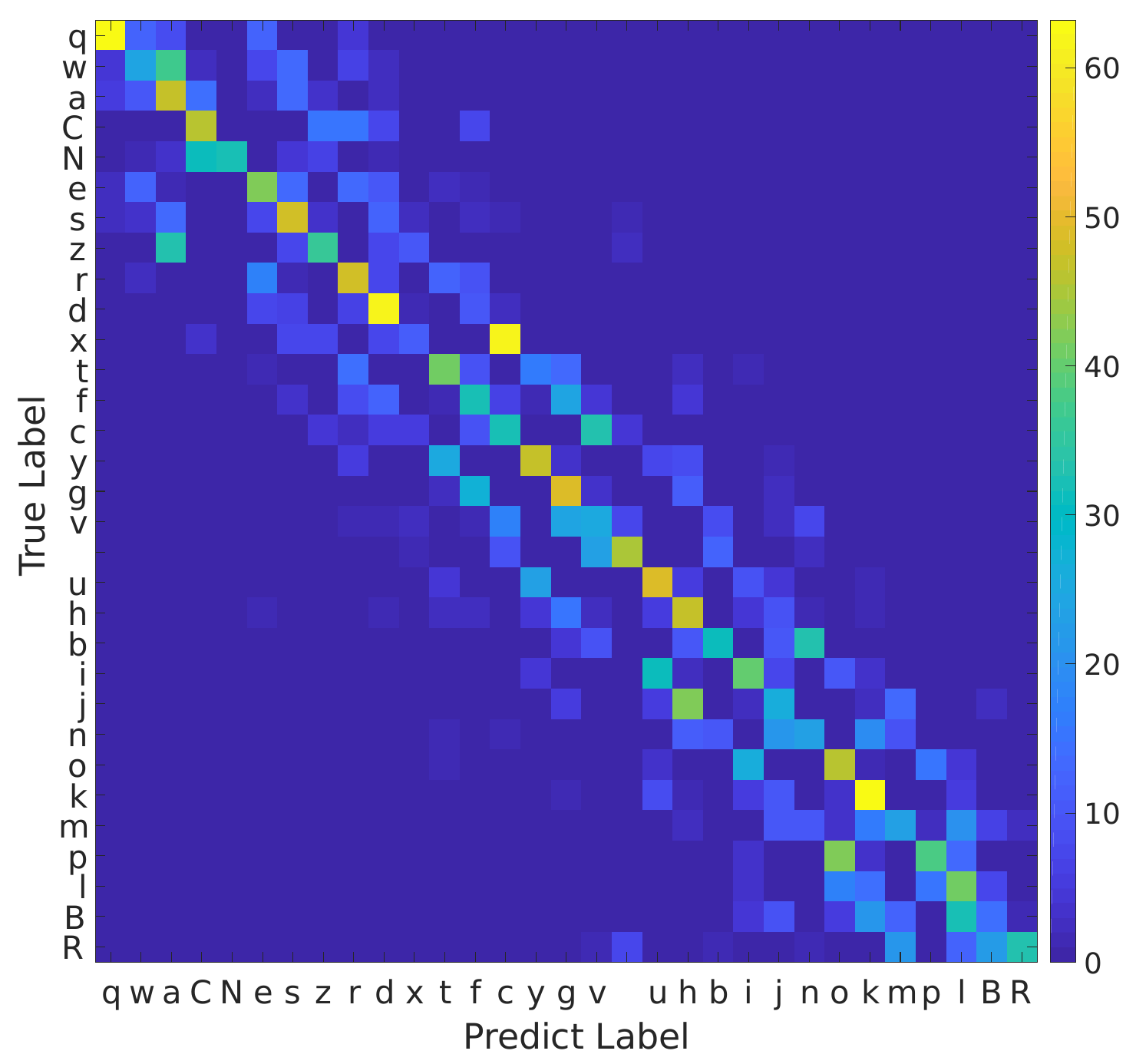}}
    \subfloat[K=1 (Numberspace)]
    {\includegraphics[width=0.45\columnwidth]{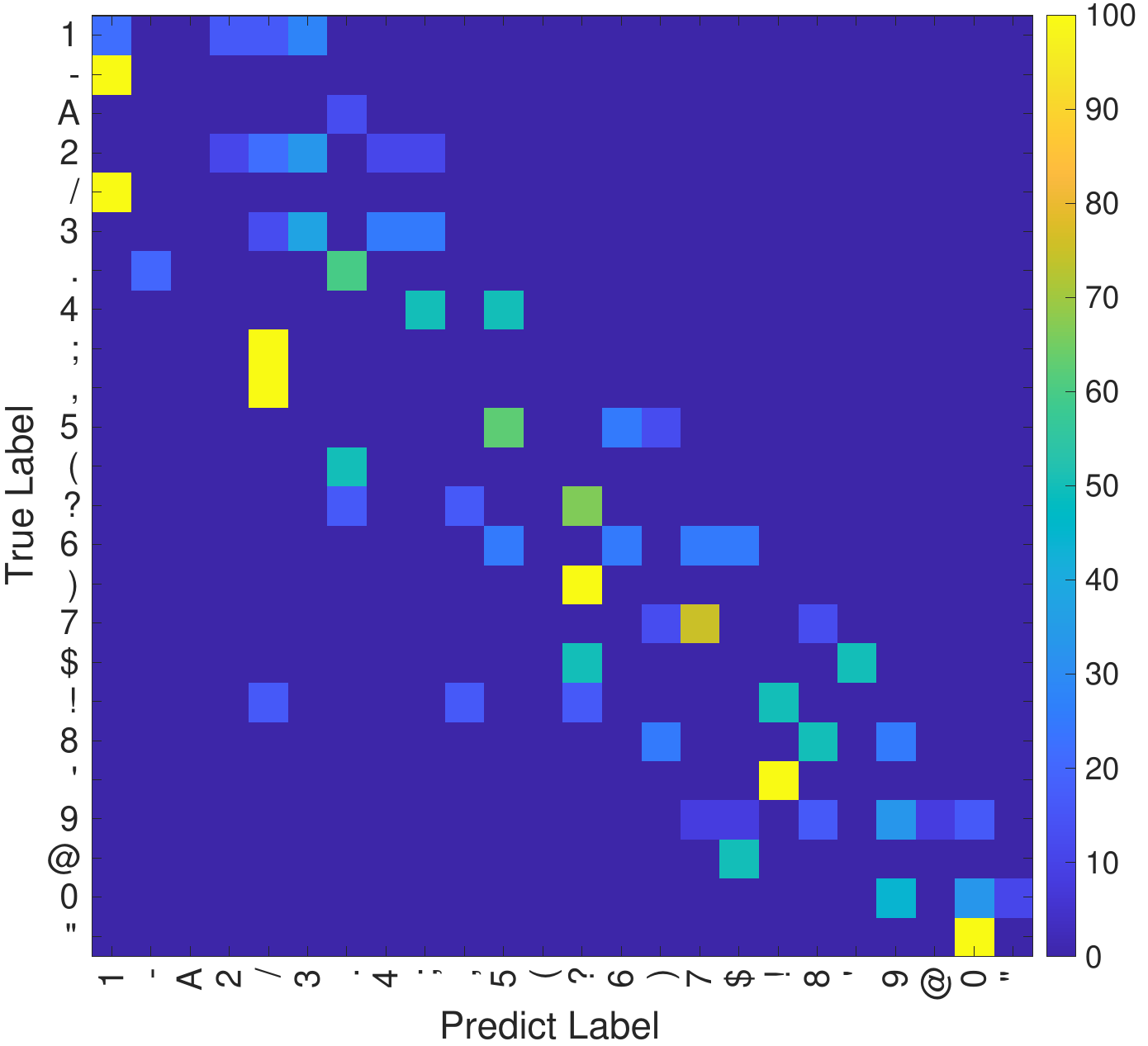}}
    \\
    \subfloat[K=5 (QWERTY)]
    {\includegraphics[width=0.45\columnwidth]{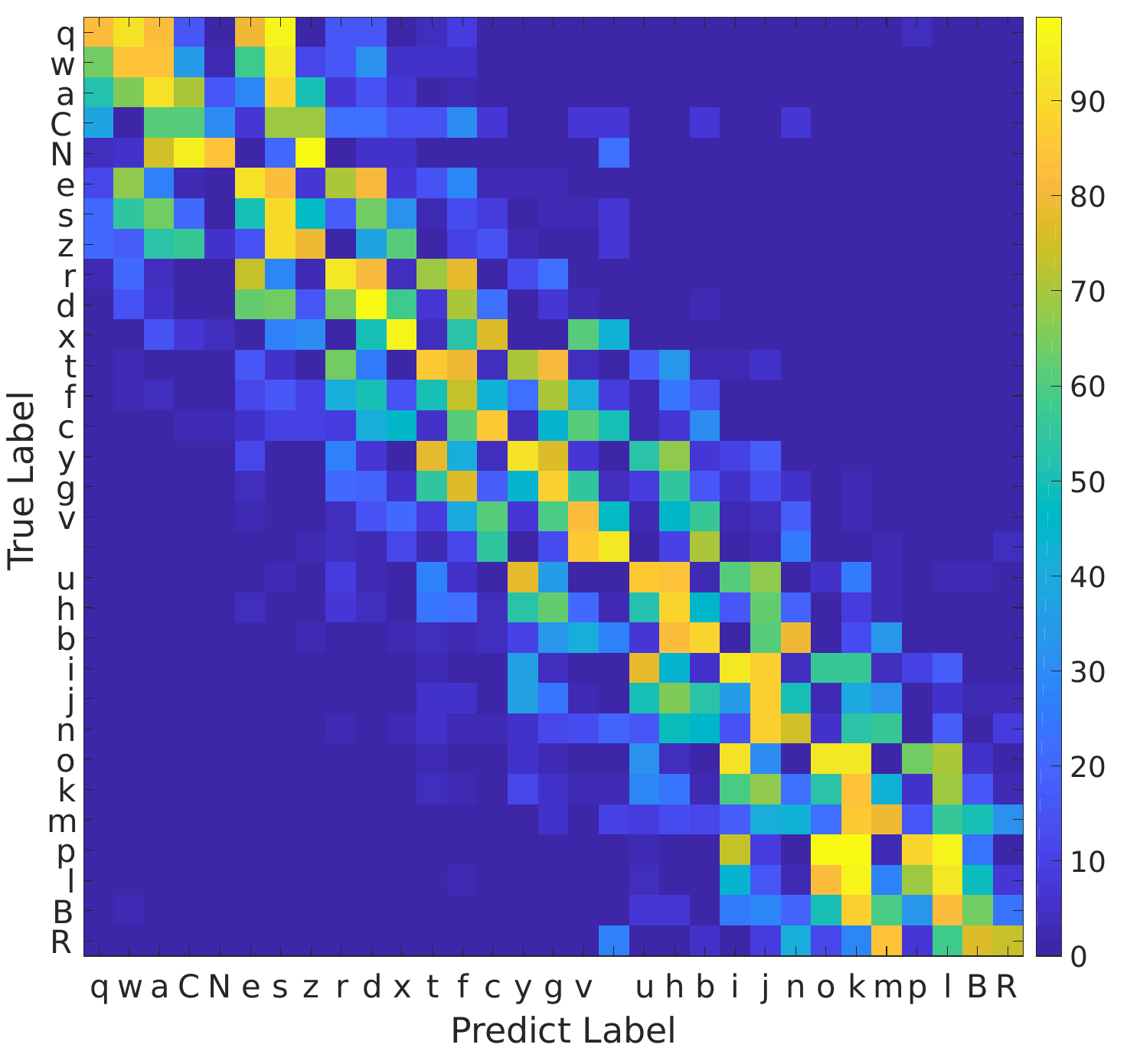}}
    \subfloat[K=5 (Numberspace)]
    {\includegraphics[width=0.45\columnwidth]{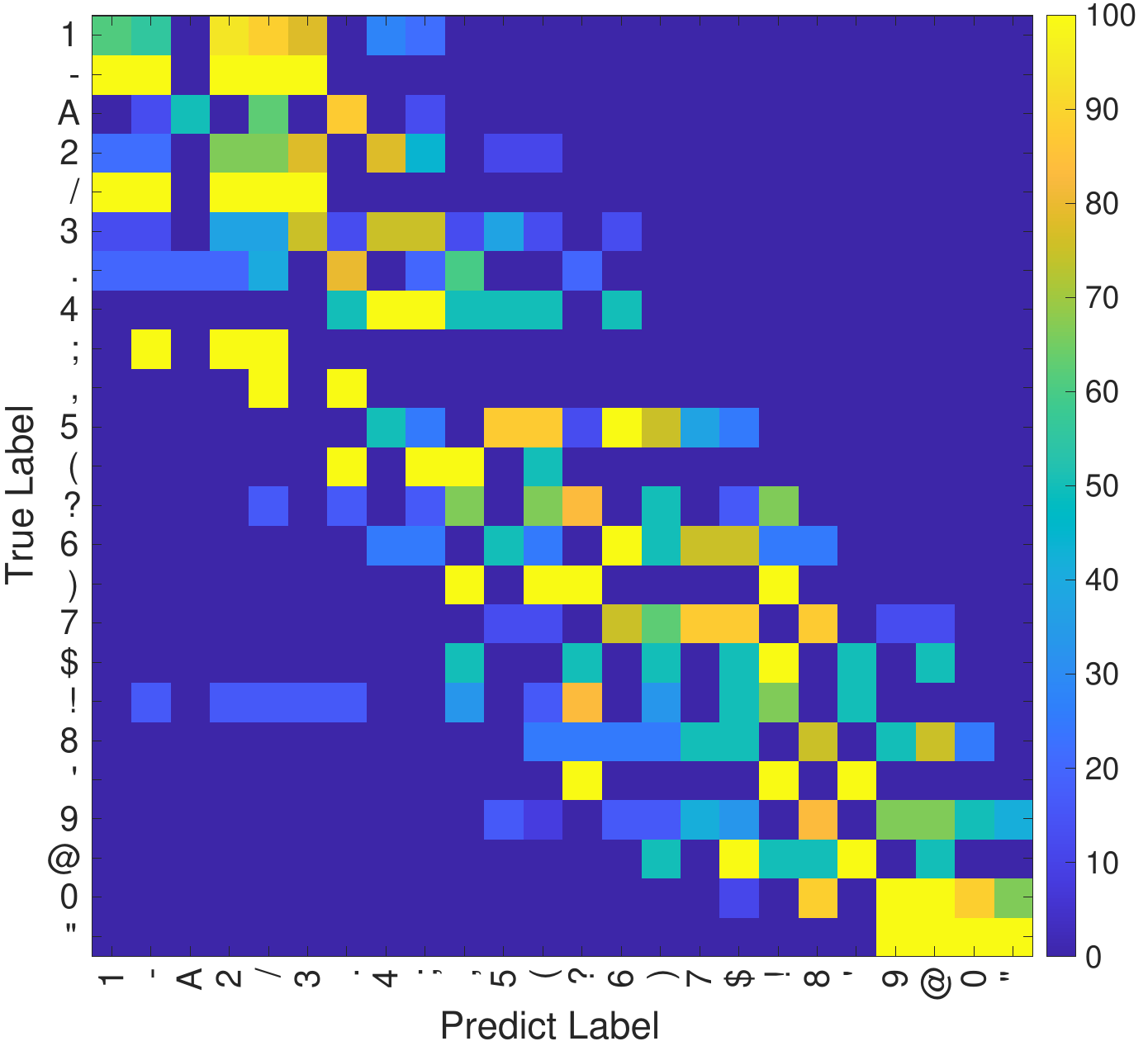}}
    \caption{confusion matrix of QWERTY / numberspace  keyboard for each character, top 1 and top 5 guesses accuracy}
    \label{fig:cmatrix}
\end{figure}

In this section, we show the confusion matrices of keystroke inference on all
characters on both the QWERTY keyboard and the numbers and special characters
keyboard in Figure~\ref{fig:cmatrix}.
From the Figure, we can clearly find the pattern that keys that are close to one
another has a higher chance of being falsely inferred.
Also, people tend to include fewer numbers and special characters. In our
dataset, the number of keystrokes on the QWERTY keyboard is 100 times the number
of keystrokes on the special characters and numbers keyboard because numbers and
special characters appear more in password entry scenarios.
Consequently, the confusion matrices for the numbers and special characters keyboard appear less smooth compared to those for the QWERTY keyboard.

Next, we show the accuracy of inferring the keystroke input of 23 participants
in different typing scenarios in Table~\ref{tab:form} including 1. message sent in social media software
2. password 3. URL and email address 4. Passcode(PIN).

\begin{table}[h]
    \centering
    \caption{Top $K$ character prediction accuracy (\%) in four typing scenarios}
    \label{tab:form}
    \begin{tabular}{|c|c|c|c|c|}
        \hline
        \textbf{$K$} & \textbf{Message} & \textbf{Password} & \textbf{URL/Email} & \textbf{Passcode} \\ \hline

        1            & 38.7             & 34.6              & 35.3               & 59.1              \\ \hline
        2            & 64.4             & 50.8              & 57.1               & 64.2              \\ \hline
        3            & 77.4             & 61.8              & 70.2               & 69.6              \\ \hline
        4            & 86.2             & 71.0              & 81.0               & 70.5              \\ \hline
        5            & 92.1             & 77.0              & 86.1               & 73.0              \\ \hline
    \end{tabular}
\end{table}

\

\noindent\textbf{Password Inference} As stated in Section~\ref{sec:3.2}, \system
attack model recognize keystrokes with letters, and numbers but without space
keys as potential passwords. Our \system model could achieve 34.6\% character
inference accuracy in the first guess and 77.0\% character inference accuracy
in the top five guesses.

According to the accuracy of correctly inferred characters, we can reduce the
password search space of a 8 to 12 character random password containing
letters and numbers from 2.2$\times$$10^{14}$$-$3.2$\times$$10^{21}$ to
            3.9$\times$$10^{5}$$-$2.4$\times$$10^{8}$. This search space decrease makes a
        brute force attack possible in seconds or hours even using an out-of-date CPU
        model like Pentium 100~\cite{mohantyintel}.

        \noindent\textbf{URL and Email address} As mentioned in
        Section~\ref{sec:3.2}, our attack recognize patterns including
        ``\texttt{@}'', ``\texttt{com}'', ``\texttt{.edu}'', ``\texttt{.us}'' as
        potential URL or email address. Our model could achieve 35.3\% character
        inference accuracy in the first guess and 86.1\% character inference
        accuracy in the top five guesses.

        \noindent\textbf{Passcode Inference} As introduced in Section~\ref{sec:3.2}, \system
        attack model recognize keystrokes restricted in a smaller typing region as
        potential passcodes. We evaluate the key pattern prediction accuracy. The first
        guess will infer 59.1$\%$ keys correctly. And with the top 5 guesses, the
        accuracy reached 73.0$\%$ as shown in Table~\ref{tab:form}. We also count how many attempts we need to recover the 6-digit passcodes. We
        could infer 8.3$\%$ of all 6-digit passcode strings in the first attempt,
        16.6$\%$ in 4 attempts, and 25.0$\%$ in 32 attempts. This success rate
        guarantees the attack model could unlock the device or hack financial apps
        within hours, which poses a great security risk.

\subsection{Message Inference}

The message inference attack is evaluated on both the messages typed in the comprehensive dataset and the additional sentence typing dataset. 
The message inference accuracy is evaluated based on three metrics: 1. character inference accuracy, 2. word segmentation accuracy, 3. word inference accuracy. 
we can tell from Table~\ref{tab:form} that the top 5 character prediction accuracy is over 92$\%$, and top 6 accuracy is 94$\%$, which does not increase significantly. Thus, we infer word segmentation and words based on the top 5 character prediction matrices.

\begin{figure}[htbp]
    \centering
    \includegraphics[width=\linewidth]{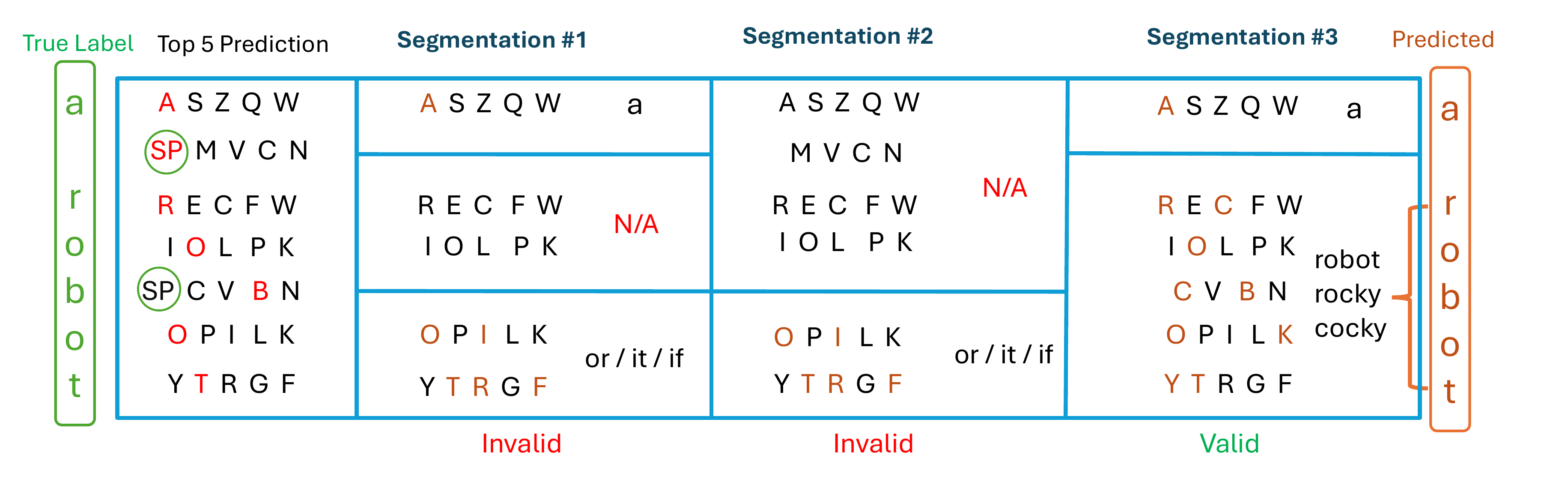}
    \caption{Segmentation Candidates Selection}
    \label{fig:segCandidate}
\end{figure}

\noindent\textbf{Sentence Segmentation} 
An example of segmenting a sentence is shown in Figure~\ref{fig:segCandidate}.
To segment sentences into words, we first identify all potential segmentation points where the ``\texttt{SPACE}'' key appears in the top 5 predictions. Treating all these keys as ``\texttt{SPACE}'' may result in false positives, as many are actually keys near ``\texttt{SPACE}''. To find correct segmentations, we iterate over possible segmentations and use a dictionary to verify potential words. We consider only segmentations that produce valid words to derive sentence candidates.
If there are no valid segmentations for more than 15 consecutive keystrokes (the largest word length in the dictionary), it indicates a true character is not within the top 5 predictions. In such cases, we skip to the next potential segment point and mark the skipped segment as unknown.

Using this method, we can significantly reduce the false positive and accurately identify ``\texttt{SPACE}'' keys to correctly segment sentences. We show the confusion matrix of ``\texttt{SPACE}'' key prediction in the sentence inference experiment in Table~\ref{tab:seg}.

\begin{table}[ht]

    \centering
    \caption{Confusion Matrix of SPACE key prediction as word segmentation in Message Inference}
    \label{tab:seg}
    \begin{tabular}{|c|c|c|}
        \hline
        \textbf{Keystrokes}   & \textbf{SPACE} & \textbf{Characters} \\ \hline
        \textbf{SPACE (Predicted)}           & 1457          & 242                     \\ \hline
        \textbf{Characters (Predicted)} & 153           & 9608                    \\ \hline
    \end{tabular}
\end{table}

\noindent\textbf{Typo Correction Inference} 
During data collection, participants are allowed to make typos and correct them using the ``\texttt{BACKSPACE}'' key. 
Such typo correction behaviors can also be identified by our method.
When a ``\texttt{BACKSPACE}'' key is in the top 5 predictions, we consider it as a potential typo correction.
If removing this key along with the key preceding it results in a valid word in the dictionary, we consider the user to have made a typo correction.
Otherwise, we ignore the ``\texttt{BACKSPACE}'' key and infer the words with the remaining letters in the top 5 predictions.
This method successfully corrected 312 out of 398 typos, with no false positives.

\begin{figure}[htbp]
    \centering
    \includegraphics[width=\linewidth]{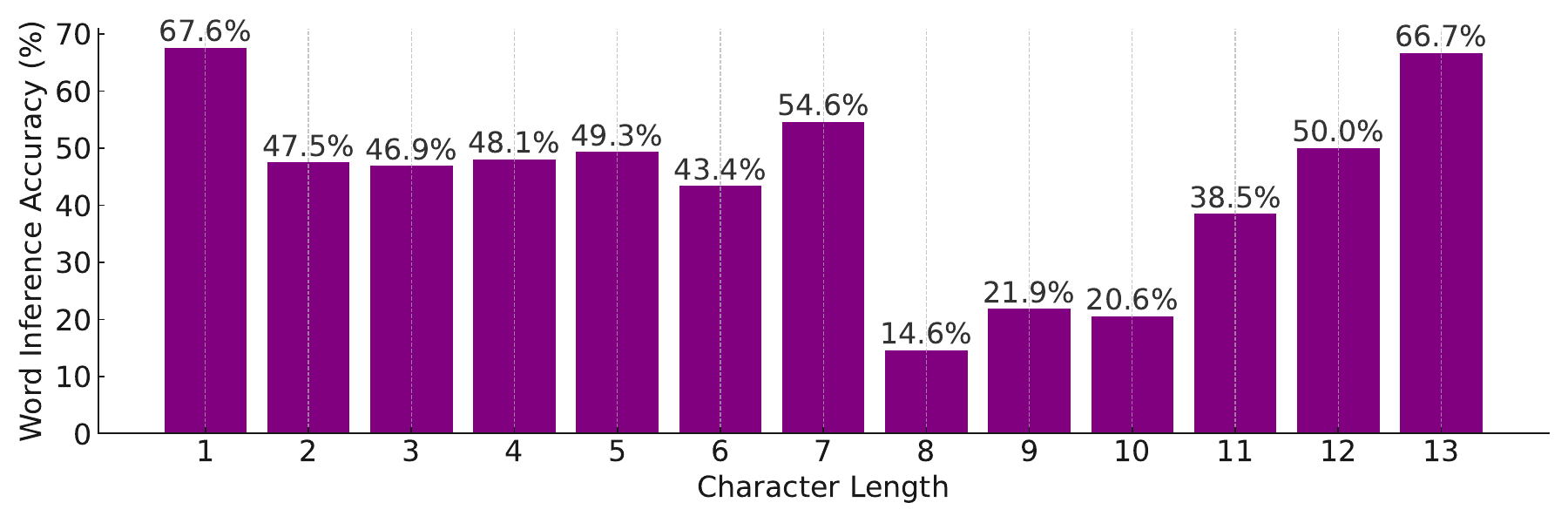}
    \caption{Word Inference Accuracy of different word length}
    \label{fig:wordCharacterLength}
\end{figure}

\noindent\textbf{Dictionary-based Word Inference} We use the 10,000 most frequent words in English from the OpenSubtitles Dataset~\cite{wictionary} for word prediction. \system attempts to infer possible words in each word segmentation based on the top 5 character predictions. The result is further filtered using the Enchant Grammar Check~\cite{enchant} to remove the non-English words. 
If multiple word candidates exist, we attempt to guess the true word in the order of the words' frequencies.
Among the 1215 correctly inferred segmentations, 858 words were identified with an average of 4.3 attempts. The word inference accuracy within 5 attempts, based on word length, is shown in Figure~\ref{fig:wordCharacterLength}. \system effectively handles words of varying lengths.%

\noindent\textbf{Consecutive Identical Keystrokes} \system has limitations in identifying multiple identical keystrokes as separate events due to the absence of saccade features, as the user's gaze does not move. This affects the inference of words like ``\texttt{better}''. 
To address this, we modified the dictionary by adding variants for words with consecutive identical keystrokes, such as ``\texttt{beter}'' for ``\texttt{better}'' and ``\texttt{wek}'' for ``\texttt{week}''. 
Using this method, 67 out of 134 words with consecutive identical keystrokes can be successfully identified.

        \subsection{\system attack Robustness}
        \label{sec:6.3}
        We have already demonstrated that the capabilities of our attack in
        remote, end-to-end, and universal scenarios. Given that the attack does
        not require any prior knowledge or recalibration on users, we evaluate
        two other aspects of attack robustness in this subsection.

        \

        \noindent\textbf{Recording Persona Virtual Camera Placement} The Persona
        virtual camera position is by default centered at the app window that is
        accessing the camera. What restricts the camera position is that the
        head pose it films cannot surpass the yaw and pitch angles up to
    $\pm$80$^{\circ}$,$\pm$80$^{\circ}$ because these angles are the max
        angles in the ETH-XGaze dataset that our gaze estimation model is
        trained on. We examined the head rotation during all typing sessions and
        find the rotations are within yaw and pitch angles
    $\pm$25$^{\circ}$,$\pm$20$^{\circ}$. This finding leads to the Persona
        virtual camera placement should be within
    $\pm$55$^{\circ}$,$\pm$60$^{\circ}$ area, centered at the direct front
of the users' head position to guarantee the gaze can be extracted
without large errors.

\begin{figure}[htbp]
    \centering
    \includegraphics[width=0.8\columnwidth]{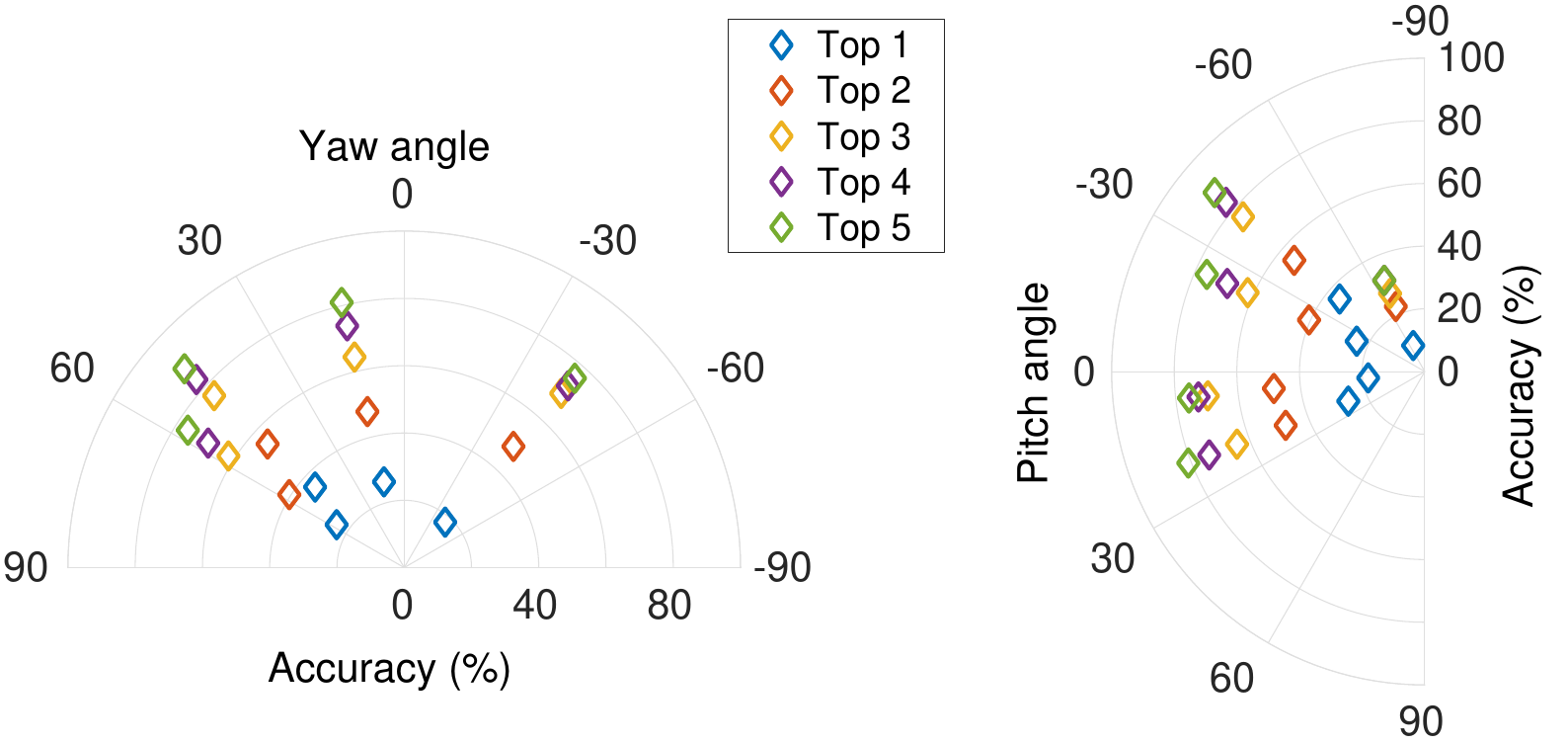}
    \caption{Key recovery accuracy with different camera placement angles}
    \label{fig:cameraPlacement}
\end{figure}

We test \system attack at these theoretical extreme virtual camera placement
angles. Our keystroke recovery accuracy does not change very much within the
angles. However, when the camera placement surpasses the angles, the accuracy
drops drastically, as illustrated in Figure~\ref{fig:cameraPlacement}.

\begin{figure}[htbp]
    \centering
    \includegraphics[width=\linewidth]{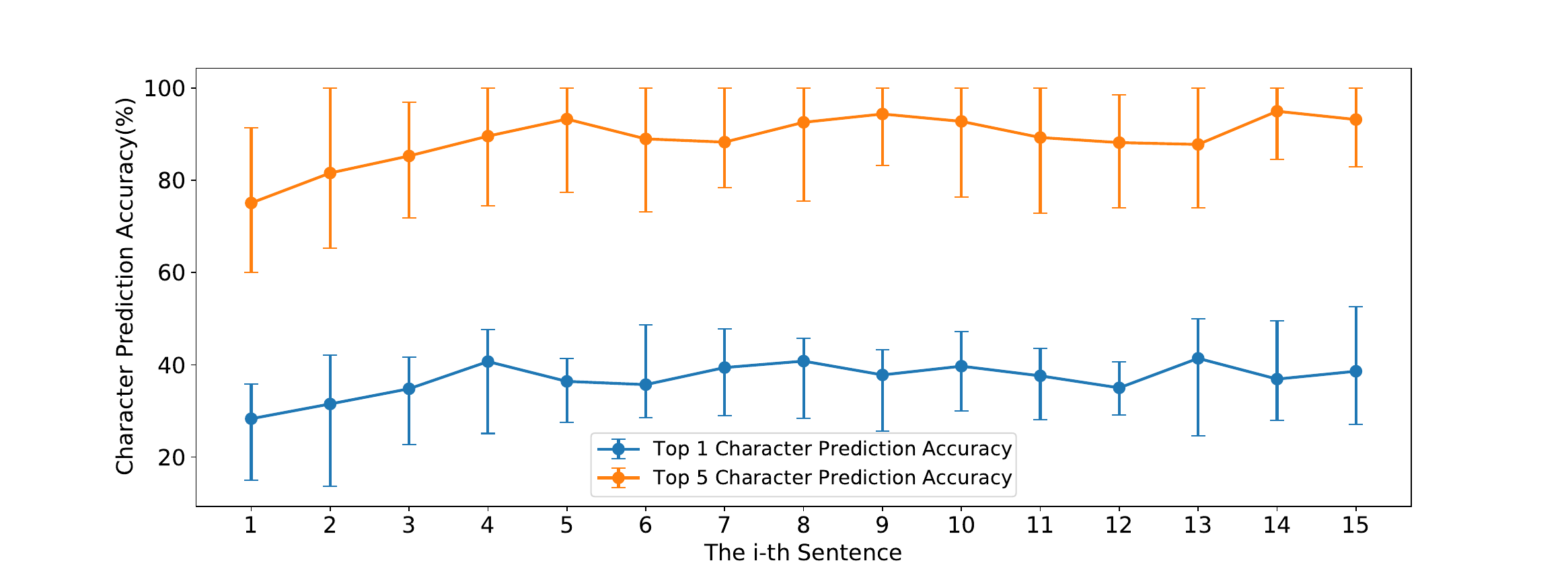}
    \caption{Character prediction Accuracy vs. 15 sentences in time sequence, averaged across participants}
    \label{fig:accuracyExperience}
\end{figure}

\noindent\textbf{Attack Performance across participants} 
We do not observe significant differences in attack performance across different genders, races, and ages. 
However, in the sentence typing inference experiment, there is a strong correlation between a participant's gaze typing proficiency and the prediction accuracy. 
To prove this, we plot the averaged character inference accuracy of all participants with deviations for the 15 sentences in Figure~\ref{fig:accuracyExperience}.
It is evident that as users gain more practice in gaze typing and become more proficient, the inference accuracy increases, especially during the initial few sentence inputs, where character prediction accuracy improves by around 10\%. 
This improvement is likely due to more proficient participants having more regular and predictable fixation and saccade patterns, as discussed in Sections~\ref{sec:4.2} and \ref{sec:4.3}. 
This observation indicates an increased threat level of our attack, as most real-world users will eventually become proficient in gaze typing.

\section{Discussion}
\label{sec:disc}

In this section, we discuss the lack of awareness about VR/MR security
vulnerabilities among hardware designers, software developers, and users, and
how this can be addressed. We also compare our work, \system, with prior works,
emphasizing the unique contributions and advantages of our method. Finally, we
propose countermeasures to mitigate the identified vulnerabilities, balancing
the need for user security with the functionality and usability of VR/MR
devices.

\subsection{Unawareness of Vulnerabilities}

There exists a common misconception among hardware designers, software
developers, and users regarding VR/MR's virtual camera. This misconception leads
them to treat it as a regular camera, overlooking the fact that it inherently
captures more biometric data due to the immersive nature of VR/MR technologies.
A notable example in our study is the eye gaze information, which can
unintentionally leak typing information.

Our analysis of applications on the Apple App Store, including popular
conference apps like Zoom, Microsoft Teams, and Slack, social media apps such as
Reddit, WeChat, Tinder, Twitter, Discord, and video chat apps like FaceTime and
Skype, reveals that many of these are vulnerable to this security flaw. This
indicates a widespread lack of awareness and understanding of these unique
vulnerabilities among developers and users alike.

However, fundamentally, hardware vendors bear the responsibility to ensure the
security of these applications. The placement of avatar views, for instance,
should not simply replicate the virtual camera's view due to the potential
exposure of sensitive data, which may be even more critical than revealing the
user's appearance itself. This reveals the need for a deeper understanding
and more careful consideration of VR/MR's unique characteristics and potential
risks among all parties involved.

\subsection{Comparison with Prior Work}

\system is an end-to-end keystroke inference attack, uniquely combining VR environments and gaze estimation, setting it apart from prior works in both domains.

\noindent\textbf{Comparison with prior work in VR} Our work, \system{}, stands out when compared with six state-of-the-art
keystroke inference attacks in VR/MR devices, as outlined in
Table~\ref{tab:comparison} detailing the practical implementation and evaluation
design.

\begin{table}[ht]
    \centering
    \caption{Comparison with prior works in VR}
    \label{tab:comparison}
    {\begin{tabular}{|>{\centering\arraybackslash}p{0.29\columnwidth}|>{\centering\arraybackslash}p{0.07\columnwidth}|>{\centering\arraybackslash}p{0.07\columnwidth}|>{\centering\arraybackslash}p{0.07\columnwidth}|>{\centering\arraybackslash}p{0.07\columnwidth}|>{\centering\arraybackslash}p{0.07\columnwidth}|}
            \hline
            \textbf{Work}                       & {\faPlane} & {\faUsers} & { \faKeyboardO} & {\faKey}     & {\faEye}   \\
            \hline
            Gear VR\cite{gearVR}                & \checkmark & \checkmark & $\times$        & \textbf{M}   & $\times$   \\
            \hline
            Typose\cite{goingMotions}           & \checkmark & $\times$   & $\times$        & \textbf{N/A} & $\times$   \\
            \hline
            Hidden Reality\cite{hiddenReality}  & $\times$   & \checkmark & $\times$        & \textbf{H}   & $\times$   \\
            \hline
            Heimdall\cite{luo2024eavesdropping} & $\times$   & \checkmark & \checkmark      & \textbf{L}   & $\times$   \\
            \hline
            VR-spy\cite{wifiVR}                 & $\times$   & \checkmark & \checkmark      & \textbf{M}   & $\times$   \\
            \hline
            \textbf{\system}                    & \checkmark & \checkmark & \checkmark      & \textbf{H}   & \checkmark \\
            \hline
        \end{tabular}}
    \\
    \faPlane: Remote Attack, \faUsers: Universal Attack, \faKeyboardO: Identify Typing Sessions, \faKey: Password Strength in Dataset, \faEye: Targeting Gaze Typing, \textbf{H/M/L}: High/Medium/Low
\end{table}

A significant advantage of \system{} is its ability to execute remote attacks by
leveraging gaze information leaked through avatars, a method that integrates
seamlessly into regular VR/MR usage. In contrast, Hidden
Reality (HR)~\cite{hiddenReality}, Heimdall~\cite{luo2024eavesdropping}, and
VR-spy~\cite{wifiVR} depend on side channels such as video, EM, and acoustics,
requiring on-site equipment like cameras or smartphones for data capture. This
makes them less feasible to carry out in practical scenarios. Meanwhile,
GearVR~\cite{gearVR} and Typose~\cite{goingMotions} utilize motion sensor data
from the headset. However, unlike \system utilizing the virtual camera in
regular VR/MR usage, these attacks require malicious apps to access the motion
sensor data which increase the complexity of the attack.

\system{} also excels in its universal applicability, a feature shared with all
other attacks except Typose which requires initial victim profiling. Regarding typing session identification, Heimdall incorporates
preliminary results analyzing click frequency to identify typing sessions and
VR-spy analyzed the frequency domain feature of channel state
information to identify typing sessions. Our
approach goes a step further by applying an RNN method to distinguish typing
sessions with high precision. This level of sophistication and accuracy is
unparalleled in other solutions.

Regarding password security, \system ensures robustness by requiring
passwords to meet Google account standards. This includes at least 8
characters involving letters, numbers, and special characters. In contrast,
GearVR and VR-spy used simpler passwords, and Heimdall opted for weak passwords
from a list of common choices. Only HR matches the complexity of
\system's password criteria, using computer-generated random passwords.

Most significantly, \system is the only solution among those compared that
targets gaze typing, an emerging feature in VR/MR technologies.
Our study found
that
users overwhelmingly prefer this intuitive and efficient input method over
hand-gesture-based or controller-based typing. Despite the growing adoption of
gaze typing in state-of-the-art VR systems and its potential to become the
standard input method, no other keystroke inference attack has considered this
feature. \system pioneers in this regard, identifying and addressing the
security vulnerabilities associated with gaze typing.

\noindent\textbf{Comparison with prior work in 2D scenarios} All works that attempt to attack tablets/smartphones/laptops without using eye gaze information are ineffective in VR gaze-typing. For instance, ``Zoom on the Keystrokes''~\cite{sabra2020zoom} focuses on shoulder movement. This feature is absent in gaze-typing on a virtual keyboard since the victim does not need to move their hands.

In contrast, several works use eye gaze information to infer keystrokes on touchscreens but face limitations when applied to VR gaze-typing. Eyetell~\cite{chen2018eyetell} utilizes features such as limbus detection which mandates fixed head postures and front-view recorded videos with narrow camera angles, making it impractical for VR scenarios. Furthermore, Eyetell identifies clicks based on \textbf{turn-points} in gaze traces, failing to recognize same-line keystrokes (e.g., ``POWER'' or ``ROUTE'') and relies on dictionary lookup, which \textbf{only works on words} but fails with more complex texts like passwords.

Similarly, GazeRevealer~\cite{wang2019your} uses the front camera of a smartphone to record the victim's view, limiting its applicability in VR where the camera can be placed at a wide-angle range. It requires the installation of a malicious app and can only infer PIN keyboards on smartphones and does not address typing session identification. While GazeRevealer employs pattern recognition to segment keystrokes, \textbf{its dataset does not include more complicated scenarios like message or password inference}.

Cazorla et al.~\cite{Cazorla2022} focus on laptops and physical keyboard and evaluate the accuracy of mapping gaze to keys using machine learning classifiers. However, they do not address keystroke segmentation or typing session identification, making their approach impractical for real-world scenarios. Additionally, they face limitations related to camera angle, which restricts their effectiveness in VR scenarios. Their work also requires participants to gaze at the physical keyboard while typing, which is not practical because experienced physical keyboard typers does not need to look at the keyboard.

\begin{table}[ht]
    \centering
    \caption{Comparison with prior works in 2D scenarios}
    \label{tab:comparison2D}
    {\begin{tabular}{|>{\centering\arraybackslash}p{0.3\columnwidth}|>{\centering\arraybackslash}p{0.08\columnwidth}|>{\centering\arraybackslash}p{0.08\columnwidth}|>{\centering\arraybackslash}p{0.1\columnwidth}|>{\centering\arraybackslash}p{0.08\columnwidth}|>{\centering\arraybackslash}p{0.08\columnwidth}|}
            \hline
            \textbf{Work} & {\faWrench} & { \faKeyboardO} & {\faKey} & {\faCamera} & {\faEye} \\
            \hline
            Eyetell\cite{chen2018eyetell} & \checkmark & $\times$ & \textbf{N/A} & \textbf{L} & $\times$ \\
            \hline
            GazeRevealer\cite{wang2019your} & $\times$ & $\times$ & \textbf{N/A} & \textbf{L} & $\times$ \\
            \hline
            Cazorla et al.\cite{Cazorla2022} & $\times$ & $\times$ & \textbf{N/A} & \textbf{L} & $\times$ \\
            \hline
            \textbf{\system} & \checkmark & \checkmark & \textbf{H} & \textbf{H} & \checkmark \\
            \hline
        \end{tabular}}
    \\
    \faWrench: Practical Attack Scenario, \faKeyboardO: Identify Typing Sessions, \faKey: Password Strength in Dataset, \faCamera: Camera Angle Range, \faEye: Targeting Gaze Typing, \textbf{H/M/L}: High/Medium/Low
\end{table}

Overall, these works demonstrate the challenges and limitations of adapting existing 2D keystroke inference techniques to VR gaze-typing scenarios, highlighting the unique strengths of \system in effectively addressing these challenges, as depicted in Table~\ref{tab:comparison2D}.

\subsection{Countermeasures}

Addressing the vulnerabilities identified in the use of Vision Pro and Persona
virtual camera involves implementing several countermeasures. These strategies
aim to balance user security and usability while mitigating risks associated
with the dual use of the eye tracker and virtual camera.

\noindent\textbf{Randomized Keyboard:} A randomized keyboard can obscure visual
feedback of keystrokes, mitigating the risk of inferring passwords or sensitive
information from eye movements~\cite{randomKeyboard}. Although this method could
effectively prevent password leakage, it may reduce typing speed and accuracy,
deviating from conventional interaction paradigms users are accustomed
to~\cite{jiang2022learning}. 

\noindent\textbf{Visual Indicators for Persona Sharing:} Persistent visual
indicators in the user's view during Persona sharing can reduce the risk of
unintentional data exposure. These include icons or overlays signaling Persona's
active status and notifications alerting users when Persona is activated. 

\noindent\textbf{Disabling Persona During Sensitive Inputs:} Similar to security
measures for screenshots during sensitive inputs, disabling the Persona view
during password or passcode entry can enhance user privacy. 
However, it may lower user experience during social
interactions using Persona.

\section{Related Work}
\label{sec:relwork}

\noindent\textbf{Keystroke Inference on VR Devices:} Various side channels can
be exploited for malicious keystroke inference during VR device usage, including
graphic-based, physical movement, acoustic, and architecture-level side
channels. GearVR~\cite{gearVR} introduced video-based and motion sensor-based
keystroke inference attacks in VR devices. Subsequent studies have explored
subtle head movements, captured by motion sensors when users type on virtual
keyboards~\cite{goingMotions, luo2022holologger} or controller movement~\cite{lee2023vrkeylogger}. Other approaches include
identifying hand gestures~\cite{hiddenReality, air-tapping} and analyzing
clicking sounds from hand controller movements~\cite{luo2024eavesdropping}.
VR-spy~\cite{wifiVR} targets EM side-channels, while Yicheng Zhang et
al.~\cite{performanceVR} target architecture-level performance data side
channels. However, none of these approaches consider gaze data, which
\system uniquely exploits for a more effective and less intrusive remote
attack.

\noindent\textbf{Video-based Keystroke Inference Attack:} Video-based keystroke
inference attacks have been extensively explored for real
keyboards and touchscreens~\cite{wang2019your, chen2018eyetell, sabra2020zoom, yang2023towards}. 
However,
these methods often impose specific conditions and assumptions that limit their
applicability. 
In contrast, \system can be
executed under a broader range of conditions. It leverages the virtual camera's
view in VR/MR systems and extract the eye gazing data, which is consistently
available and unaffected by physical world constraints.

\noindent\textbf{Appearance-based Gaze Estimation:} Zhang et al. proposed a CNN
and appearance-based gaze estimation~\cite{zhang2015appearance} and published a
large dataset~\cite{zhang2017mpiigaze} for gaze estimation research. They later
published the ETH-XGaze dataset~\cite{zhang2020eth}, which includes over 1
million images of participants gazing at a wide range. \system uses a model
trained on this dataset, leveraging its robustness for effective gaze-based
keystroke inference.

\section{Conclusion}
\label{sec:conclu}
In this paper, we introduced \system, a keystroke inference attack leveraging
eye gaze data from VR/MR devices. Our research highlighted significant security
vulnerabilities with eye-tracking technologies. Through experiments, we demonstrated how attackers could
reverse engineer confidential keystrokes by analyzing video recordings of eye
movements during text entry. Apple has implemented robust security measures to mitigate these risks. Future efforts
should also focus on 
improving security protocols to protect users in VR/MR
environments and 
exploring other attacks exploiting the leaked gaze information
in VR/MR devices.

\section{Acknoledgement}
\label{sec:ack}
This work was partially supported by the National Science Foundation under Award Number 2028897 and 1916175 and partially supported by Intel Gift Donation.

\bibliographystyle{ACM-Reference-Format}
\bibliography{vr}

\appendix

\end{document}